\PassOptionsToPackage{pdfpagelabels=false}{hyperref}

\documentclass[fleqn,usenatbib,useAMS]{mnras}

\usepackage{graphicx}	% Including figure files
\usepackage{amsmath}	% Advanced maths commands
\usepackage{amssymb}	% Extra maths symbols
\usepackage{color}
\usepackage[figure,figure*,table,table*]{hypcap}

\input{macros.sty}

\title[Brightest galaxies as halo centre tracers]
      {Brightest galaxies as halo centre tracers in SDSS DR7}
\author[J.~U.~Lange et al.]
{Johannes~U.~Lange$^{1, 2}$,
	Frank~C.~van~den~Bosch$^1$,
	Andrew~Hearin$^3$,\newauthor
	Duncan~Campbell$^1$,
	Andrew~R.~Zentner$^4$,
	Antonio~Villarreal$^4$, and\newauthor
	Yao-Yuan~Mao$^4$\\
	$^{1}$Department of Astronomy, Yale University, P.O. Box 208101, New Haven, 
	CT\\
	$^{2}$Kavli Institute for Theoretical Physics, University of California, 
	Santa Barbara, CA\\
	$^{3}$Yale Center for Astronomy \& Astrophysics, Yale University, New 
	Haven, CT\\
	$^{4}$Department of Physics and Astronomy \& Pittsburgh Particle Physics, 
	Astrophysics, and Cosmology Center (PITT PACC),\\
	\phantom{$^{4}$}University of Pittsburgh, Pittsburgh, PA 15260\\}

\pubyear{2017}

\begin{document}

\date{Accepted xxx. Received xxx}

\label{firstpage}
\pagerange{\pageref{firstpage}--\pageref{lastpage}}

\maketitle

\begin{abstract}
  Determining the positions of halo centres in large-scale structure
  surveys is crucial for many cosmological studies. A common
  assumption is that halo centres correspond to the location of
  their brightest member galaxies. In this paper, we study the
  dynamics of brightest galaxies with respect to other halo members in
  the \textit{Sloan Digital Sky Survey} DR7. Specifically, we look at
  the line-of-sight velocity and spatial offsets between brightest
  galaxies and their neighbours. We compare those to detailed mock
  catalogues, constructed from high-resolution, dark-matter-only
  $N$-body simulations, in which it is assumed that satellite galaxies trace
  dark matter subhaloes. This allows us to place constraints on the
  fraction $f_{\rm BNC}$ of haloes in which the brightest galaxy is
  not the central. Compared to previous studies we explicitly take
  into account the unrelaxed state of the host haloes, velocity
  offsets of halo cores and correlations between $f_{\rm BNC}$ and the
  satellite occupation. We find that $f_{\rm BNC}$ strongly decreases
  with the luminosity of the brightest galaxy and increases with the
  mass of the host halo. Overall, in the halo mass range $10^{13} - 10^{14.5}
  \Msunh$ we find $f_{\rm BNC} \sim 30\%$, in good agreement
  with a previous study by Skibba et al. We discuss the implications of
  these findings for studies inferring the galaxy--halo connection from
  satellite kinematics, models of the conditional luminosity function
  and galaxy formation in general.
\end{abstract}

\begin{keywords}
methods: statistical -- galaxies: kinematics and dynamics -- galaxies: groups: 
general -- cosmology: dark matter
\end{keywords}

\section{Introduction}

In the standard cosmological model, cold dark matter (CDM) haloes provide the 
potential wells that enable the condensation of gas into stars. Ultimately, 
this causes CDM halos to be the natural cites for the formation of galaxies 
\citep[see][for a review]{MBW}. Thus, galaxies and the dark matter haloes in 
which they reside constitute an inseparable entity. 

Hierarchical structure formation is a fundamental prediction of the 
$\Lambda$CDM cosmological model. In this paradigm, small dark matter haloes 
form first and grow by the accretion of surrounding dark matter mass 
and other smaller haloes. Accreted dark matter haloes can survive 
inside the the {\em host} halo for many dynamical times. These accreted 
haloes are called {\em subhaloes} and they are associated with {\em satellite} 
galaxies. The galaxy harboured by the host halo is known as the {\em central} 
galaxy. 

It is generally assumed that satellite galaxies accreted by a bigger 
halo are slowly quenched by processes unique to the cluster or group
environment and stop forming new stars, whereas the central galaxy
continues to grow. Additionally, once satellite galaxies are disrupted
by gravitational interactions inside the host, it is assumed that part
of their stellar material is accreted onto the central galaxy. Given
this picture, it is expected that central galaxies form a distinctive
population of galaxies that are systematically more massive and more
luminous than their satellite galaxies. In fact, it is commonly
assumed that the brightest galaxy inside a dark matter halo is often,
if not always, the central galaxy.

Indeed, observations have shown that brightest halo galaxies (BHGs)
form a distinctive population of halo members. Their luminosities
cannot be predicted by defining them as the brightest galaxy out of a
population whose luminosities are drawn from the same underlying
distribution \citep[e.g.,][]{Tremaine+77, Lin+10, Hearin+13, Shen+14}. This can 
be interpreted as BHGs being more luminous {\it because} they are central
galaxies of the halo. However, there is no firm reason to assume
that BHGs {\em always} coincide with the centres of dark matter halos. 
For example, central galaxies could have spatial or
velocity offsets from the actual halo centre. Additionally, it is
expected that in a fraction $f_{\mathrm{BNC}}$ of the haloes the
brightest (or most massive) galaxy is {\em not} the central, but a satellite
galaxy. This has been analysed specifically for galaxy clusters
\citep[see, e.g.][]{Sanderson+09, Zitrin+12, Hikage+13, Wang+14, Lauer+14,
Hoshino+15} for which the large number of satellite galaxies and
the bright X-ray emitting gas make it relatively easy to determine the
halo centre. However, there are so far very few studies analysing the
phase-space positions of BHGs in low-mass haloes.

Recently, \cite{Guo+15b} modelled the redshift-space clustering of galaxies in 
SDSS allowing centrals to have velocity offsets with respect to their dark 
matter haloes.  Their results suggest that even in low-mass haloes central 
galaxies are not at rest relative to the dark matter. \cite{vdBosch+05}
and \cite{Skibba+11} have analysed the phase-space coordinates of BHGs
in SDSS, concluding that central galaxies are either not at rest with
respect to the dark matter halo or some satellites are BHGs, i.e. $f_{\rm
BNC} \neq 0$. More specifically, \cite{Skibba+11} provided evidence
that the latter case is the main reason for the mis-centering of
BHGs. They analysed $f_{\rm BNC}$ for a large range of halo masses
down to $\sim 10^{12.5} \ h^{-1} M_\odot$ and found values of
$f_{\mathrm{BNC}}$ that increase from $\sim 25\%$ for Milky-Way
size haloes to $\sim 45\%$ for massive clusters. Interestingly, those
values are higher than predicted by semi-analytical models (SAMs) of
galaxy formation \citep{Skibba+11}. The authors
argued that their finding could be interpreted as evidence for longer
quenching and dynamical friction time-scales than commonly assumed in
galaxy formation theory. However, the results of \cite{Skibba+11} are
in tension with those of \cite{Hoshino+15}, who find a
significantly lower $f_{\rm BNC}$ of $\sim 20 - 30\%$ for clusters.

In this work, we employ an analysis technique similar to the one used
by \cite{Skibba+11}. Specifically, we analyse offsets in phase-space
between BHGs and other halo members. We improve upon the analysis
of \cite{Skibba+11} by taking into account several effects that could
potentially bias the inferred values of $f_{\rm BNC}$. These effects include velocity offsets of halo cores and central galaxies with respect to the main halo, halo triaxiality, the unrelaxed state of the satellite population and the correlation of $f_{\rm BNC}$ with satellite occupation. Additionally, we
characterize how $f_{\rm BNC}$ depends on {\it both} halo mass $M$ and
BHG luminosity $L_{\rm BHG}$ instead of halo mass alone. The dependence on $L_{\rm BHG}$ is particularly important because it is observationally more easily accessible than halo mass.

Determining $f_{\mathrm{BNC}}$ is crucial for many cosmological
studies. For example, weak lensing surveys often explore gravitational
lensing around brightest cluster galaxies (BCGs) \cite[see,
e.g.][]{Sheldon+09a, Sheldon+09b} or satellites \cite[see,
e.g.][]{Yang+06, Li+13, Li+14, Sifon+15, Li+16, Niemiec+17}. Knowing 
$f_{\mathrm{BNC}}$ is necessary in this case to get an unbiased estimate of the
corresponding dark matter halo mass \citep{Johnston+07, Li+14}. Additionally,
modelling the (redshift-dependent) clustering of galaxies using halo
occupation distribution (HOD) modelling implicitly requires knowledge
of $f_{\rm BNC}$. For example, the average satellite occupation of
dark matter halos is typically scaled by the average central
occupation \citep{Zheng+07, Zehavi+11, Leauthaud+12}. Alternatively,
assuming that only haloes that host a central galaxy can also host
satellites above a given threshold \citep[see, e.g.][]{Guo+15b} is equivalent to
assuming that $f_{\rm BNC} = 0$. Furthermore, many studies of the kinematics of
satellite galaxies implicitly assume $f_{\rm BNC} =
0$ \citep{vdBosch+04, More+09a, More+09b, More+11, Li+12} or that
satellite galaxies cannot be brighter than $x$ times the central where
$x$ is some number above unity \citep{Klypin+09, Wojtak+13}. In studies
using satellite kinematics to constrain the
galaxy--halo connection, a non-zero $f_{\mathrm{BNC}}$ would lead to a
systematic over-prediction of halo masses \citep{Skibba+11}. Furthermore,
\cite{Tinker+17} have recently shown that determining central galaxies in
galaxy groups is crucial for studying galaxy assembly bias \citep{Zentner+14, 
Zentner+16} and galactic conformity \citep{Weinmann+06, Kauffmann+13}. Finally,
determining $f_{\mathrm{BNC}}$ can inform our understanding of galaxy
formation and evolution since it depends on many poorly
constrained processes, such as satellite quenching, dynamical
friction, and the degree to which central galaxies grow in mass by
accreting satellites. 

In this work, we find values for $f_{\rm BNC}(M, L_{\rm BHG})$ that are 
surprisingly  large compared to theoretical models and predictions from 
conditional luminosity functions (CLFs) \citep{Yang+03, vdBosch+03}. 
Marginalized over the luminosity dependence, we find $f_{\rm BNC}(M)$ values 
that broadly agree with \cite{Skibba+11}. We show explicitly what our best-fit 
values for $f_{\rm BNC}$ imply for studies of satellite kinematics, the 
radial profile of satellite galaxies, and galaxy-galaxy lensing. We also find that standard CLF models 
fail to predict the observed $f_{\rm BNC}(M, L_{\rm BHG})$ and describe how 
those models could be adjusted to be in better agreement with observations. 
Additionally, we discuss how our results fit into galaxy formation models.

Our paper is organized as follows. In section \ref{sec:Methods} we
outline how we measure the phase-space positions of BHGs with respect
to nearby galaxies, primarily other halo members. In
section \ref{sec:Obervations} we describe how we extract the signal
from the main galaxy-sample of SDSS DR7 \citep{Abazajian+09}. Next, we
present how we produce detailed mock galaxy catalogues with a
parametrized $f_{\rm BNC}$ and compare it to observations in
section \ref{sec:Mock Catalogues}. Our main results are described in
section \ref{sec:Results} and we explore the implications in
section \ref{sec:Discussion}. Finally, our findings are summarized in
section \ref{sec:Conclusion}. Throughout this work, we adopt a flat
$\Lambda$CDM cosmology with $\Omega_m = 0.307$, $n = 0.96$ and
$\sigma_8 = 0.8228$. Relevant quantities are expressed in units of $h = H / 
(100 \ \mathrm{km/s/Mpc})$.

\section{Methods}
\label{sec:Methods}

In this section we describe our general method to detect the fraction
of haloes in which the BHG is not the central.

\subsection{Summary Statistics}

To analyse $f_{\rm BNC}$ we use the method introduced
by \cite{vdBosch+05} and also used in \cite{Skibba+11}. The idea is to
detect spatial and velocity offsets of the BHG with respect to other
halo members. In a hypothetical, fully-relaxed halo containing an unlimited 
number of satellite galaxies, the average phase-space
position of all galaxies in the halo should coincide with the halo
centre. Thus, by comparing the position of the BHG to the galaxies in
the same halo we can determine whether or not the BHG is the central
galaxy. In actuality, the number of galaxies detectable in each halo is
finite and halos are not fully relaxed, so the centre of an individual halo 
cannot be determined. Rather, we can determine $f_{\rm BNC}$ for a large 
ensemble of dark matter haloes.

\cite{vdBosch+05} introduced the
$\mathcal{R}$-statistic which is designed to detect line-of-sight
velocity offsets of the BHG with respect to neighbouring galaxies.
In the following, we will refer to BHGs as primaries and their nearby
galaxies as secondaries. The $\mathcal{R}$-statistic is defined for each
primary as follows,
\begin{equation}
\mathcal{R} = \sqrt{N_{\rm scd}} \, \frac{\langle v_{\rm scd} \rangle - v_{\rm pri}}{\hat{\sigma}_{\rm scd}},
\end{equation}
where $N_{\rm scd}$ is the number of secondaries likely associated with
the same underlying halo as the primary, $\langle v_{\rm
scd} \rangle - v _{\rm pri}$ is the average line-of-sight velocity of
the secondaries with respect to the primary,
\begin{equation}
\langle v_{\rm scd} \rangle = \frac{1}{N_{\rm scd}} \sum\limits_{i = 1}^{N_{\rm scd}} v_{i},
\end{equation}
and $\hat{\sigma}_{\rm scd}$ is an estimate of the velocity dispersion
for each system,
\begin{equation}
\hat{\sigma}_{\rm scd}^2 = \frac{1}{N_{\rm scd} - 1} \sum\limits_{i = 
1}^{N_{\rm scd}} \left( v_{i} - \langle v_{\rm scd} \rangle \right)^2.
\end{equation}

\begin{figure}
  \centering \includegraphics[width=\columnwidth]{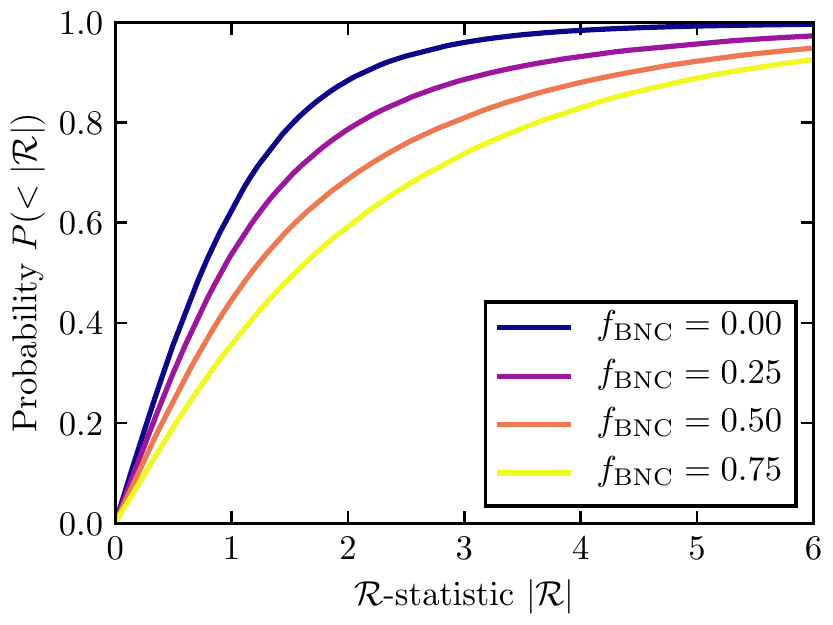} \caption{Demonstration
  of the sensitivity of the $\mathcal{R}$-statistic to the value of
  $f_{\rm BNC}$. We show the cumulative distribution of
  $|\mathcal{R}|$ for different values of $f_{\rm BNC}$, as indicated. We assume a
  fixed number of secondaries, $N_{\rm scd} = 5$, and that
  satellite velocities are drawn from a Gaussian distribution while
  the central is at rest.}  \label{fig:R_Statistic}
\end{figure}

If all secondaries were true satellites and all $v_{i}$ were drawn from a 
Gaussian distribution, the $\mathcal{R}$-statistic would 
follow a student's t-distribution. On the other hand, if some of the brightest 
galaxies were satellites, the distribution would be biased towards higher
values in $|\mathcal{R}|$.

Figure \ref{fig:R_Statistic} shows the cumulative distribution of 
$|\mathcal{R}|$ for different values of $f_{\rm BNC}$. We take the absolute 
value of $\mathcal{R}$ because there is no useful information in the sign of
$\mathcal{R}$. For this plot we make several idealized
assumptions. For example, we assume that the number of secondaries
is fixed at $N_{\rm scd} = 5$ and that the line-of-sight velocities of
satellites are drawn from a single Gaussian distribution. For $f_{\rm
BNC} = 0$, the distribution equals a student's t-distribution with
virtually no values of $|\mathcal{R}|$ exceeding $6$. As we increase
$f_{\rm BNC}$, higher values in $|\mathcal{R}|$ become more
likely. This is the general effect we will utilize throughout this
study. It is important to note that many of the assumptions going into
Figure \ref{fig:R_Statistic} are unlikely to be met in
observations. For example, in general $N_{\rm scd}$ will vary
from primary to primary, and the velocity dispersion of satellites
likely has a radial profile $\sigma(r)$. In addition,
there might be velocity anisotropy and/or non-vanishing
higher-order moments of the velocity
distribution \citep[e.g.,][]{Diemand+04, Wojtak+13}. Thus, the line-of-sight
velocities of satellite galaxies cannot be drawn from a single
Gaussian distribution. Furthermore, it is not possible to identify
all halo members while having no contamination from interloper
galaxies. In observations some true satellite galaxies will be missed,
while some other secondaries will be interloper galaxies belonging to a
different halo. Finally, effects of the survey mask and spectroscopic
fibre collisions can further alter the
$\mathcal{R}$-distribution. Thus, it is necessary to compare
observational values for the $\mathcal{R}$-distribution to detailed
mock catalogues that take all of the above effects into
account \citep{vdBosch+05, Skibba+11}.

As the $\mathcal{R}$-statistic is designed to detect line-of-sight
velocity offsets, we similarly define the $\mathcal{S}$-statistic to
detect spatial offsets perpendicular to the line-of-sight:
\begin{equation}
\mathcal{S} = \sqrt{N_{\rm scd}} \, \frac{\sqrt{\langle x_{\rm scd} \rangle^2 + \langle y_{\rm scd} \rangle^2}}{\sqrt{\hat{\sigma}_{x, \rm scd}^2 + \hat{\sigma}_{y, \rm scd}^2}},
\end{equation}
where $x$ and $y$ represent physical distances with respect to
the primary perpendicular to the line-of-sight and the average and
dispersion estimates are defined in analogy with the
$\mathcal{R}$-statistic. \cite{Skibba+11} have shown that using the
$\mathcal{R}$ and $\mathcal{S}$-statistic simultaneously helps to
break model degeneracies. Specifically, it helps to distinguish
between a scenario where $f_{\rm BNC} > 0$ and a scenario in which the
centrals have spatial and velocity offsets with respect to the centre
of the potential well of the halo. In principle, both hypotheses
affect the $\mathcal{R}$ and $\mathcal{S}$ distributions in a similar
way by driving them towards more extreme values. However, if the
central has spatial and velocity offsets it will have a stronger
effect on the $\mathcal{R}$-statistic than on the
$\mathcal{S}$-statistic because the central galaxy would need large velocity 
offsets before it is able to escape the centre of the potential well. 
\cite{Skibba+11} showed that offsets
between the central galaxy and the potential well cannot
simultaneously explain both the $\mathcal{R}$ and $\mathcal{S}$
distributions in SDSS data, and therefore argued for $f_{\rm BNC} >
0$.

\subsection{Isolation Criteria}

The $\mathcal{R}$ and $\mathcal{S}$ statistics require the identification of 
primaries and secondaries. Ideally, all BHGs should be identified as primaries 
while all other galaxies that reside in the same halo are identified as 
secondaries. \cite{vdBosch+05} and \cite{Skibba+11} used a group finder for 
this task. The brightest galaxy
in each group was chosen to be the primary and all remaining galaxies
in that group associated secondaries. The advantage of this method
is that all galaxies in the survey can be used. However, this type of
completeness comes with a number of possible failure modes.
\cite{Campbell+15} identified two main group finding errors. Galaxies that are 
part of a single large halo may be erroneously identified as two distinct 
groups, a process called ``fracturing,'' while galaxies of two distinct haloes 
may erroneously be associated with a single group, an error referred
to as ``fusing.'' The former case will lead to at least one primary not being a 
BHG and the latter one will result in secondaries that are not part of the same 
halo as the primary. \cite{Campbell+15} showed that around $30\%$ of all
supposed satellites, i.e. secondaries, identified by the \cite{Yang+07} group 
finder are actually centrals of their own, i.e. mostly BHGs. Group 
identification errors are likely to influence the 
measured $\mathcal{R}$ and $\mathcal{S}$ distributions. If the incidence of 
these group errors is not reproduced in the mock catalogues used to interpret 
the data, group errors may lead to biases in the inferred values of $f_{\rm 
BNC}$.

To mitigate this potential bias we use the isolation
criteria introduced in \cite{vdBosch+04}. Generally, this method uses
a cylindrical isolation criteria. A galaxy is considered a primary if
it is not within a cylindrical volume defined by depth $(\Delta V)_\rmh$ and radius $R_\rmh$ of another galaxy brighter by a factor $f_\rmh$. The cylinder is aligned with the line of sight to account for redshift space distortions. Having identified isolated galaxies as primaries, we associate secondaries to them if they lie in a smaller cylinder defined by $(\Delta V)_\rms$ and radius $R_\rms$ and are fainter by a factor $f_\rms$. We follow \cite{vdBosch+04} and
adopt $f_\rms = f_\rmh = 1$. 

Additionally, \cite{vdBosch+04} have
shown that one can achieve a very high purity and completeness in the
satellites by making the cylinder sizes dependent on the luminosity
of the galaxy. The idea is that the luminosity of the galaxy scales
with the halo mass which in turn scales with the size of the halo.
\cite{More+11} found that the average velocity dispersion as a function
 of the luminosity of the primary scales as
\begin{equation}
\log \sigma_{200} = -0.11 + 0.37 \log L_{10} + 0.30 (\log L_{10})^2\,.
\label{eq:cylinder}
\end{equation}
Here $\sigma_{200}$ is the satellite velocity dispersion in units of
$200 \kms$, and $L_{10} = L / (10^{10} \Lsunh)$.

For this work we adopt $(\Delta V)_\rms = (\Delta V)_\rmh = 5 \sigma =
1000 \, \sigma_{200} \kms$, $R_\rms = 0.25 \, \sigma_{200} \, \Mpch$ and
$R_\rmh = 3.2 \, R_\rms$.  Since Eq.~(\ref{eq:cylinder}) implies
unrealistically high values for the satellite velocity dispersion for
a few extremely bright galaxies, specifically those without
spectroscopic redshift, we cap the satellite
velocity dispersion to $1000 \kms$ when specifying our cylinders.
These values are chosen as a compromise between completeness
and contamination. Altogether, we can achieve interloper
fractions of around $18\%$, as measured in our mock catalogues described in 
section \ref{sec:Mock Catalogues}. In other words, $82\%$ of all secondaries
are galaxies living in the same halo as the primary. Similarly, 
in a mock with $f_{\rm BNC} = 0$, we find
that more than $99\%$ of all primaries are true BHGs. 

We note that in this method, some galaxies are classified as neither primaries
nor secondaries. In principle, this does not pose a problem because the $\mathcal{R}$ and
$\mathcal{S}$-statistics do not require all galaxies in the same halo
to be identified. However, we point out an asymmetry in satellite identification: 
completeness in true satellites is typically close to unity in the line-of-sight direction around primaries, 
but incompleteness is larger in the transverse direction because galaxies beyond a certain projected distance
are not included in the calculation. 
This incompleteness reduces the sensitivity of the $\mathcal{S}$-statistic but 
does not bias our estimator. 

\section{Observations}
\label{sec:Obervations}

\begin{figure*}
  \centering \includegraphics[width=\textwidth]{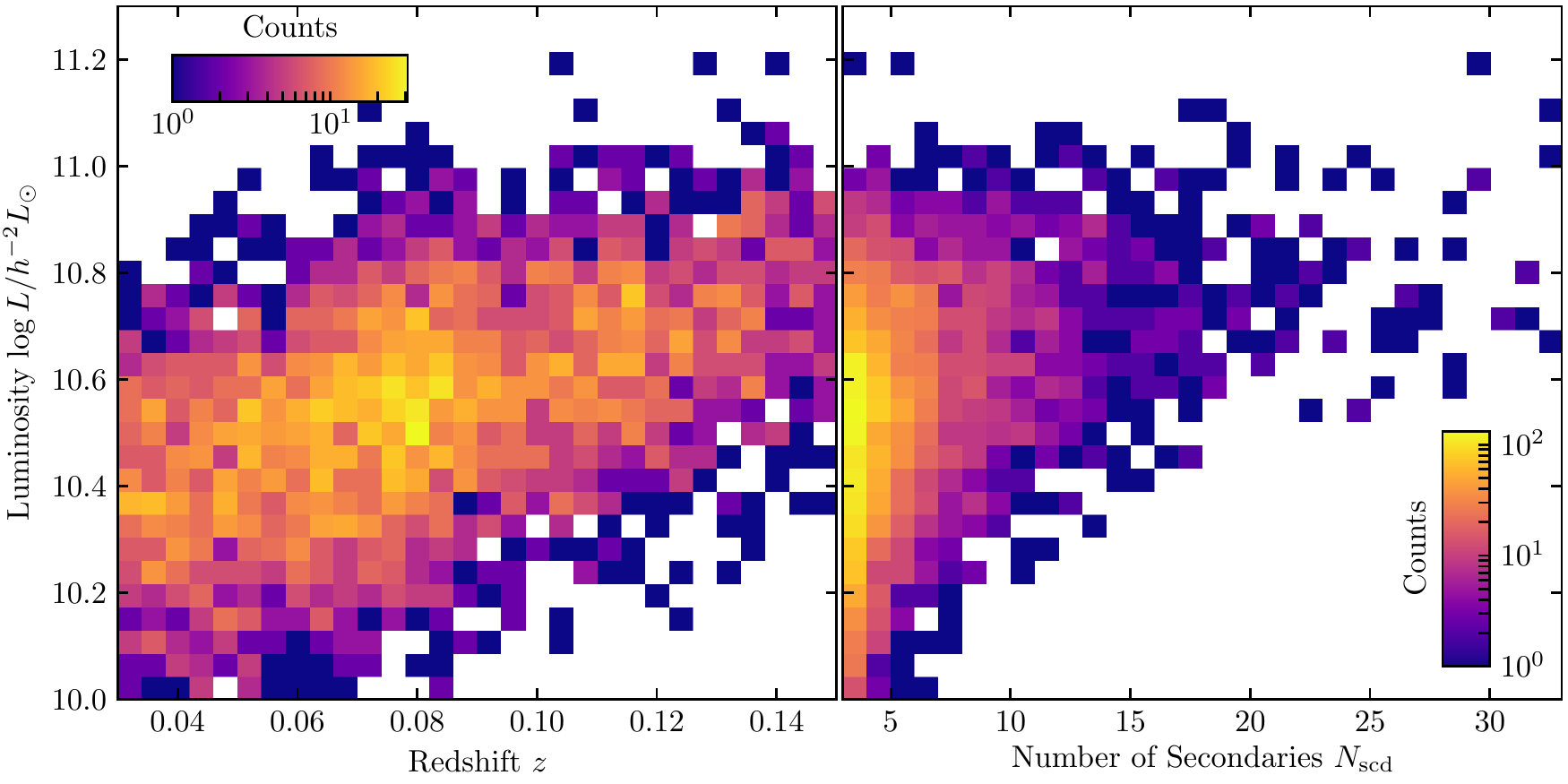} \label{fig:Primaries} \caption{Properties
  of primaries in SDSS DR7. The left-hand panel shows a histogram of
  the luminosities of the primaries against their redshifts. Due to the
  flux limit of SDSS, there is a positive correlation of redshift with
  luminosity. The right-hand panel is similar with the x-axis being
  the number of associated secondaries. Most primaries
  have less than $10$ secondaries.}
\end{figure*}

In this work, we use data from the SDSS DR7 main galaxy
sample \citep{Padmanabhan+08, Abazajian+09}. In particular, we work with
the \texttt{bright0} sample of the New York University Value-Added
Galaxy Catalog \citep[NYU-VAGC;][]{Blanton+05}. This catalogue
includes all galaxies of the SDSS main galaxy sample with an apparent
magnitude $m_r < 17.6$. Due to the design of the SDSS survey, if two
galaxies are separated by less than $55''$, spectroscopic fibres
cannot be placed for both of them. Due to these fibre collisions a
small fraction of the galaxies lack spectroscopic redshifts. In
the \texttt{bright0} sample, galaxies lacking a spectroscopic redshift
have been assigned the redshift of the closest neighbour. We convert
apparent $r$-band magnitudes into luminosities
k-corrected \citep{Blanton+07} and evolution corrected to $z =
0.1$. Of the \texttt{bright0} sample we select those galaxies with
$0.01 \leq z \leq 0.17$ and an r-band luminosity $L > 10^9 \Lsunh$. 
Roughly $500,000$ of the $570,000$ galaxies in the original
sample pass this cut. The redshift and luminosity limits are
determined primarily by the volume and resolution of our mock
catalogues, as discussed in the next section.

As the next step we run the isolation criteria discussed in the
previous section on this sample. After having identified primaries and
their associated secondaries we apply additional selection cuts. We
require that all primaries have a spectroscopic redshift, i.e. no
fibre collision, and lie in a region of SDSS DR7 with at least $80\%$
redshift completeness. We stress that we apply the isolation criteria
using all galaxies, i.e. also those with fibre collisions. In that case we use 
the redshift of the closest neighbour as given in the \texttt{bright0} sample. 
The reason is that the true BHG might itself not have a spectroscopic
redshift. Excluding this galaxy before applying the isolation criteria
would result in the second brightest galaxy of the halo being wrongly
identified as a primary. Next, we require the primary to have a
redshift in the range $0.03 \leq z \leq 0.15$. This redshift cut is
more restrictive than the previous one and designed to avoid edge
effects in redshift space. This more restrictive cut is not applied to
associated secondaries around that primary. Particularly, the redshift range 
for secondaries is larger by $0.02$, corresponding to $\sim 6000 \kms$, which 
is larger than the maximum $(\Delta V)_\rms$ of $5000 \kms$. Furthermore, we 
exclude galaxies
that lie close to the survey edge. We quantify this by calculating an
angular completeness score for every galaxy. The completeness score is
defined as the fraction of the area defined by the cylinder given by
$R_\rmh$ around each galaxy that lies within the window function of
the \texttt{bright0} sample. We require a completeness score of at
least $90\%$ for primaries. Again, this cut is not applied
to associated secondaries. We demand that primaries have
$L \geq 10^{10} \Lsunh$ and at least $3$ secondaries with
spectroscopic redshifts. Finally, following \cite{Skibba+11}, we require 
$\hat{\sigma}_{\rm scd} \geq 50 \, \kms$.

Upon applying these selections, we arrive at a sample of roughly $3600$ 
primaries and calculate the $\mathcal{R}$ and $\mathcal{S}$-statistics for each 
primary using only secondaries with spectroscopic redshifts. We show the 
luminosity and redshift distributions of the primaries in Figure 
\ref{fig:Primaries}. As seen in the left-hand panel, there is a general trend 
of increasing luminosity with redshift that is due to the flux limit of SDSS 
DR7. In the same Figure we also show the distribution of the number of 
secondaries around each primary. We find that most primaries have $3$ to
$10$ secondaries. Also, there are more secondaries around
brighter primaries. Altogether, we find around $20,000$ secondaries
in our sample.

\section{Mock Catalogues}
\label{sec:Mock Catalogues}

In this section we describe how we create mock catalogues for
parameter inference. Generally, we create mock galaxy catalogues
similar to the \texttt{bright0} sample of the NYU-VAGC and then
process them using the same pipeline as for the real data.

\subsection{Dark Matter Simulation}

In this work we use results from the Small MultiDark
Planck \citep[SMDPL;][]{Klypin+16} dark-matter-only $N$-body
simulation. The simulation traces the evolution of dark matter
structures in a cubic volume of $400 \Mpch$ on a side
using $3840^3$ particles, resulting in a particle mass of $9.6 \times
10^7 \Msunh$. The cosmological parameters of this simulation
are compatible with results from the \textit{Planck}
satellite \citep{Planck14}. The simulation output has been analysed
with the ROCKSTAR halo finder \citep{Behroozi+13} from which we use
the redshift $z = 0.1$ halo catalogue\footnote{\url{http://yun.ucsc.edu/sims/SMDPL/hlists/index.html}}. This redshift roughly corresponds to the 
median redshift of primaries in our SDSS sample. Furthermore, we require dark 
matter (sub)haloes in the halo catalogue to have $M_{\rm peak}$, the maximum 
halo mass achieved over the lifetime of the (sub)halo, exceeding $\sim 3 
\times 10^{10} \ \Msunh$, which is $300$ times the dark matter particle 
mass. This ensures that all (sub)haloes in the catalogue are well resolved.

\subsection{Galaxy Occupation}

\begin{table}
	\centering
	\begin{tabular}{c | c}
		Parameter & Value\\
		\hline\hline
		$\log L_0 / (\Lsunh)$ & $9.95$\\
		$\log M_1 / (\Msunh)$ & $11.24$\\
		$\sigma_{\log L}$ & $0.157$\\
		$\gamma_1$ & $3.18$\\
		$\gamma_2$ & $0.245$\\
		$\alpha_\rms$ & $-1.18$\\
		$b_0$ & $-1.17$\\
		$b_1$ & $1.53$\\
		$b_2$ & $-0.217$\\
		\hline
	\end{tabular}
	\caption{Parameters of the default CLF parametrization used in this work.}
	\label{tab:CLF_default_parameters}
\end{table}

We use \texttt{halotools} \citep{Hearin+16} to populate dark matter
haloes from the SMDPL simulation output with galaxies. In particular,
we use a CLF \citep{Yang+03} approach, the parametrization of \cite{Cacciato+09} and the best-fit \texttt{Fiducial} parameters from \cite{Cacciato+13}. These parameters are listed in Table \ref{tab:CLF_default_parameters}.

The occupation of dark matter haloes with galaxies is
split into a central and satellite galaxy part,
\begin{equation}
\Phi (L|M) = \Phi_\rmc (L|M) + \Phi_\rms (L|M),
\end{equation}
where $L$ is the luminosity of the galaxy (in units of $\Lsunh$) and $M$ the mass of the dark
matter halo (in units of $\Msunh$). Following \cite{Cacciato+13}, the dark matter halo mass
is taken to be the mass defined with respect to $200$ times the
background density of the Universe, $M_{200b}$. The median luminosity
of central galaxies is parametrized by
\begin{equation}
\langle \log L_\rmc (M) \rangle = \log \left( L_0 \frac{(M / M_1)^{\gamma_1}}{\left[ 1 + (M / M_1) \right]^{\gamma_1 - \gamma_2}} \right),
\end{equation}
where $L_0$, $M_1$, $\gamma_1$ and $\gamma_2$ are parameters taken
from \cite{Cacciato+13}, see Table~\ref{tab:CLF_default_parameters}. Each halo 
from the halo catalogue is assigned
a central galaxy and luminosities are drawn from a log-normal
distribution with median $\log L_\rmc (M)$ and scatter $\sigma_{\log L} =
0.157$, i.e.
\begin{equation}
\Phi_\rmc(L|M) = \frac{\log e}{\sqrt{2\pi\sigma_{\log L}^2} L} \exp\left[ - 
\frac{(\log L - \langle \log L_\rmc (M) \rangle)^2}{2 \sigma_{\log 
L}^2} 
\right].
\end{equation}
The satellite CLF is parametrized by
\begin{equation}
\Phi_\rms (L|M) = \frac{\Phi_\rms^\star(M)}{L_\star} \left( \frac{L}{L_\star} \right)^{\alpha_\rms} \exp\left[ - \left( \frac{L}{L_\star} \right)^2 \right]
\label{eq:CLF_sats}
\end{equation}
with
\begin{equation}
\log \Phi_\rms^\star(M) = b_0 + b_1 \, \log M_{12} + b_2 \, (\log M_{12})^2
\end{equation}
and
\begin{equation}
\log L_\star(M) = \langle \log L_\rmc(M) \rangle - 0.25,
\label{eq:Cut-off_luminosity}
\end{equation}
which describes the exponential cut-off of the satellite luminosity
function. Here $M_{12} = M/(10^{12}\Msunh)$ and 
$\alpha_\rms$, $b_0$, $b_1$ and $b_2$ are parameters (see 
Table~\ref{tab:CLF_default_parameters}) again
taken from \cite{Cacciato+13}. The number of satellites is determined
by first integrating Eq.~(\ref{eq:CLF_sats}) over the relevant
luminosity range. This gives the expected number of satellites. We
then assume that the actual number follows a Poisson
distribution. Note that for the moment we assume the faint-end slope
$\alpha_\rms$ of the satellite luminosity function to be independent
of the halo mass. Also, it is noteworthy that the best-fit parameters
of \cite{Cacciato+13} have been derived for different cosmological
parameters and partially using data-sets different from SDSS DR7. We
will explore the robustness of our results to reasonable changes in
the above choice of parameters in section \ref{subsec:CLF_Dependence}.

Finally, we stress that for the moment we allow satellites to be
brighter than their respective centrals. In fact, for a given CLF
parametrization, the fraction of haloes of mass $M$ and with a
brightest halo galaxy of luminosity $L_{\rm BHG}$ in which this
brightest galaxy is not the central, which we denote by $f_{\rm BNC}(M,L_{\rm BHG})$, 
is fully determined. It can be expressed via
\begin{equation}
f_{\rm BNC}(L_{\rm BHG}, M) = \frac{\Phi_\rms(L_{\rm BHG}|M)}{\frac{\Phi_\rmc(L_{\rm BHG}|M)}{1 - \int\limits_{L_{\rm BHG}}^\infty \Phi_\rmc(L|M) \rmd L} + \Phi_\rms(L_{\rm BHG}|M)}.
\label{eq:f_BNC_CLF}
\end{equation}
The first term in the denominator accounts for the possibility that
the brightest galaxy is the central, while the second term and the
term in the numerator represent the case of the satellite being the
brightest galaxy. Note that the first term is not just
$\Phi_\rmc(L_{\rm BHG}|M)$, but boosted by $\left[1 - \int_{L_{\rm
BHG}}^\infty \Phi_\rmc(L|M) \rmd L\right]^{-1}$. The reason is that we assume
that all haloes have exactly one central which increases $\Phi_\rmc$
once we know that no central is brighter than $L_{\rm BHG}$. This and other 
useful relations for CLF models are described in the appendix. We will
later swap phase-space positions of centrals and satellites to
re-parametrize $f_{\rm BNC}$, as described in section 
\ref{subsec:f_BNC_parameters}. Finally, note that $f_{\rm BNC}$ as predicted by 
the CLF model depends sensitively on the exponential cut-off of the satellite 
luminosity function given in Eq. (\ref{eq:CLF_sats}) and 
(\ref{eq:Cut-off_luminosity}) \citep{Skibba+11}. In section 
\ref{subsec:CLF_Implications}, we will test the impact of variations to the 
above parametrization on the prediction for $f_{\rm BNC}$.

\subsection{Phase-Space Coordinates}

\subsubsection{Central galaxy velocities}
\label{subsubsec:central_velocities}

The phase-space coordinates of central galaxies are derived
from the properties of their host haloes. It has been shown that
central galaxies are unlikely to be at rest with respect to the bulk velocity of
the entire halo. For example, \cite{Guo+15a, Guo+15b}, analysing the
redshift-space clustering of galaxies in the SDSS DR7 and BOSS survey,
have argued that central galaxies must have additional velocity
offsets of the order of $25\%$ of the dark matter particle velocity
dispersion of the host halo. Additionally, the authors have shown that
the strength of these velocity offsets increases with halo
mass. Similar conclusions were reached by \cite{Lauer+14}. Intriguingly, this 
kind of velocity offset is similar to that of halo cores with respect to the 
bulk velocity seen in dark matter simulations \citep{Behroozi+13}. Indeed, 
further studies have shown that central galaxies trace the dark matter halo 
core much better than the centre-of-mass of the entire halo \citep{Guo+16, 
Ye+17}. Because \cite{vdBosch+05} have shown that such velocity offsets of the 
central from the overall halo can have considerable impact on the 
$\mathcal{R}$-statistic, it is important to take this effect into account when 
modelling $f_{\rm BNC}$. Deriving $f_{\rm BNC}$ without it, as done in 
\cite{Skibba+11}, could lead to an overestimation of $f_{\rm BNC}$. In this 
work, the position of the central coincides with the dark matter density peak 
of the host halo and the velocity is set to the dark matter core velocity. The 
latter is defined as the center-of-mass velocity of the particles that enclose 
the inner 10 percent of the host halo radius \citep{Behroozi+13}.
\begin{figure}
  \centering \includegraphics[width=\columnwidth]{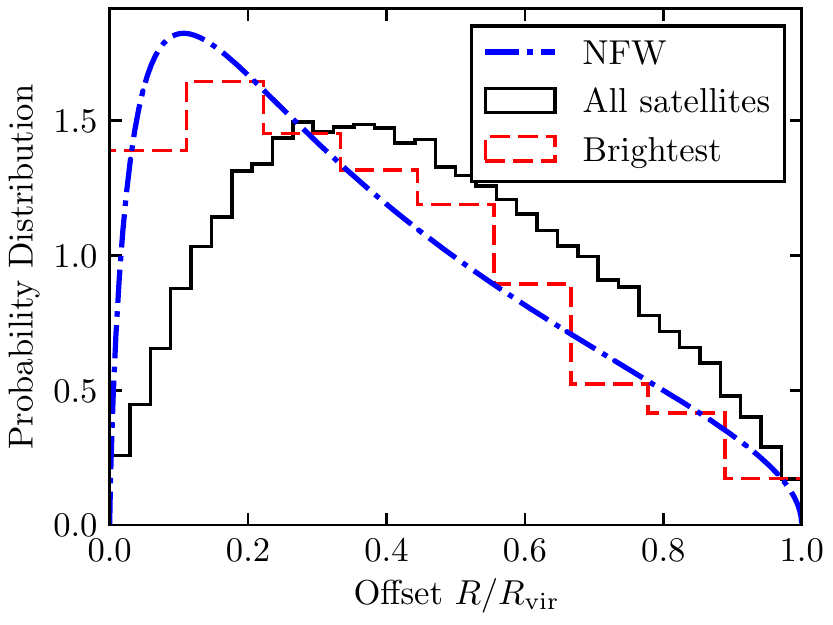} \caption{Comparison
  of the spatial distributions of all satellite galaxies (black,
  solid) against the brightest satellites (red, dashed) and the dark
  matter distribution (blue, dot-dashed) in our mock galaxy
  catalogues. In all haloes, the phase-space position of the brightest 
  satellite 
  is that of the subhalo with the highest $M_{\rm peak}$. The dark matter 
  distribution is modelled with an NFW profile where the concentration 
  parameter equals the median of all host halos. Our 
  model implies that satellite galaxies are anti-biased with respect to the 
  dark matter and that the brightest satellite galaxies are more centrally 
  concentrated than the entire satellite galaxy population.}  
  \label{fig:Distribution_Satellites}
\end{figure}

\subsubsection{Satellite galaxy velocities}
\label{subsubsec:satellite_velocities}

Satellite galaxy phase-space positions are assigned based on the
phase-space positions of actual subhaloes inside $R_{\rm vir}$. For
every halo in the simulation we first draw satellite galaxy numbers
and luminosities from the CLF assuming a Poisson distribution. We then
rank-order the satellite galaxies in each halo according to their
luminosity. Next, we assign phase-space positions of subhaloes in the
same halo rank-ordered by their $M_{\rm peak}$ value. Note that the number of 
satellites is drawn independently of the number of resolved subhaloes
that are present inside a host halo. Consequently, it is possible that the 
number of satellite galaxies assigned to a halo exceeds the number of 
subhaloes. If that occurs, we take relative phase-space positions from random 
subhaloes hosted by haloes of a similar mass and apply a random angular 
orientation. However, even for the lowest luminosity threshold used in
this analysis, this only needs to be done for $\sim 5\%$ of all the
satellite galaxies.

This novel approach combines the flexibility of
the traditional CLF method with the predictive power of subhalo
abundance matching (SHAM) \citep[e.g.,][]{ValeOstriker04, Conroy+06, Guo+10, Hearin+13b}. 
Because subhaloes are spatially anti-biased
with respect to the dark matter distribution \citep{Diemand+04}, our
model implies the same for satellite galaxies. Additionally, since we
assume a direct correlation between $M_{\rm peak}$ and $L$, any
segregation with respect to $M_{\rm peak}$ \citep{vdBosch+16}
manifests itself as a luminosity segregation. In particular our
approach implies that the brightest satellite galaxies are radially
more concentrated than the overall satellite galaxy population. This
is demonstrated in Figure \ref{fig:Distribution_Satellites} where we
show the projected core separations for satellite galaxies above a
luminosity threshold of $10^9 L_\odot$ in haloes of $10^{14} \Msunh \leq M_{\rm vir} 
\leq 3 \times 10^{14} \Msunh$ as the solid, black histogram. 
This should be compared to both the distribution of core 
separations of the brightest satellite galaxy (red, dashed 
histogram) and the dark matter distribution described by a 
Navarro--Frenk--White 
\citep[NFW,][]{Navarro+96} profile (blue, dash-dot line) with a concentration 
parameter set to $5.4$, the  median concentration of all host halos. These 
results are in good qualitative agreement with the observational results of 
\cite{Hoshino+15}.

\subsubsection{Correlated satellite velocities}
\label{subsubsec:sat_vel_correlations}

As mentioned in \cite{Skibba+11}, satellite velocities might be
correlated among each other. For example, two recently merged haloes
might be severely unrelaxed, resulting in strong phase-space
correlations of the satellite population. Also, the filamentary
structure of the cosmic web might introduce similar
effects. \cite{Skibba+11} have shown that ignoring such correlations
can bias the inferred value of $f_{\rm BNC}$ high. Because they did
not take this effect into account, they regarded their derived values
for $f_{\rm BNC}$ as upper limits. Our approach for assigning
phase-space coordinates of satellites naturally accounts for such
correlations. 

Let us first explore the presence of satellite velocity correlations. To quantify this effect, 
we compute
\begin{equation}
\mathcal{C} = \sqrt{\left\langle \frac{\langle (v_i - v_\rmc) (v_j - v_\rmc) \rangle_{\rm sub}}{V_{\rm vir}^2} \right\rangle},
\end{equation}
where $v_i$ and $v_j$ are the $z$-velocities of distinct subhaloes,
$v_\rmc$ is the velocity of the halo core, $\langle \rangle$
denotes the average over all haloes and $\langle \rangle_{\rm sub}$
the average over all subhaloes of a particular halo. Ideally, because satellite 
galaxies sample the entire halo, and not merely
the core region, we would want to express subhalo velocities relative
to the bulk velocity of the halo, rather than its core
velocity. However, these phase-space positions are not available in
the publicly available halo catalogues. As a consequence, we
already expect $\mathcal{C} > 0$, simply due to the velocity offsets
of halo cores. In order to account for this, we proceed as follows.
Let us assume that there are no correlations between subhalo and core
velocities in the rest-frame of the entire halo, so that
$\langle v_i v_j \rangle_{\rm sub} = 0$ for $i \neq j$ and $\langle
v_i v_\rmc \rangle_{\rm sub} = 0$. Then we find
\begin{equation}
\mathcal{C} = \sqrt{\left\langle \frac{v_\rmc^2}{V^2_{\rm vir}} \right\rangle}.
\end{equation}
Hence, if the above assumption of no satellite correlation is true,
$\mathcal{C}$ will be the same, but not necessarily zero, regardless of the subhalo population
used. We can assess this by splitting the subhaloes in `young' and
`old' sub-populations according to $M_{\rm vir} > M_{\rm peak} / 3$
and $M_{\rm vir} \leq M_{\rm peak} / 3$, respectively\footnote{Here
the notion is that old subhaloes were accreted earlier, and have
therefore experienced more mass loss \citep{vdBosch+05b, Zentner+05}.}. We compute $\mathcal{C}$ over
all haloes with $13 \leq \log M_{\rm vir} / (\Msunh) \leq 14$
that have at least $3$ young and $3$ old subhaloes. Unbound subhaloes
and higher-order subhaloes (i.e., sub-subhaloes, etc.) have been
excluded for this analysis. We find $\mathcal{C} = 0.171 \pm 0.003$
and $\mathcal{C} = 0.090 \pm 0.004$ for young and old subhaloes,
respectively. On the other  hand, using all subhaloes we find $\mathcal{C} = 
0.122 \pm 0.004$. Again, if the non-vanishing $\mathcal{C}$ would be
entirely due to core velocity offsets, those values should be
identical. Instead, the data suggests that there are additional
correlations in the velocities of recently accreted subhaloes, i.e. $\langle 
v_i v_j \rangle_{\rm sub} > 0$.

\subsubsection{Summary}
\label{subsubsec:velocity_summary}

To summarize, we assign phase-space positions for central and satellite 
galaxies using the phase-space positions of halo centres and subhaloes in the 
simulation from which the dark matter halo catalogs are generated. In this way, 
we account for the unrelaxed state of the haloes, including halo core velocity 
offsets, subhalo velocity correlations, and asphericity. Most of these effects
increase the absolute values of the $\mathcal{R}$ and
$\mathcal{S}$-statistics. Thus, taking it into account is important to
get an unbiased estimate of $f_{\rm BNC}$.

\subsection{Parametrizing \texorpdfstring{$f_{\rm BNC}$}{f\_BNC}}
\label{subsec:f_BNC_parameters}

So far, our choice of the CLF model implicitly determines $f_{\rm
BNC}$ as a function of halo mass and luminosity of the BHG 
(Eq.~[\ref{eq:f_BNC_CLF}]). However, as shown in section 
\ref{subsec:CLF_Implications}, the best-fit $f_{\rm BNC}$ implied by the data 
is considerably different than the CLF prediction. Since
the main goal of this work is to model $f_{\rm BNC}$ regardless of the
detailed CLF choice, we re-parametrize $f_{\rm BNC}$. First, for each
halo in the catalogue we swap the luminosities of the central and the
brightest satellite galaxy if the brightest satellite is brighter than
the central. This implies $f_{\rm BNC} = 0$ for all haloes. This
procedure is similar to the one in \cite{Skibba+11}. At this
point, the question is how to model the possibility that some
satellite galaxies are brighter than the central. \cite{Skibba+11}
assumed that this probability is dependent only on the mass of the
dark matter halo. Generally, this probability should also depend on
the luminosity of the BHG. This did not affect the results
of \cite{Skibba+11} since the BHG luminosity was not used in their
analysis. However, because we explicitly use $L_{\rm BHG}$ as an
observable, we need to parametrize $f_{\rm BNC}$ as a function of both
halo mass and BHG luminosity.

\begin{figure*}
  \centering \includegraphics[width=\textwidth]{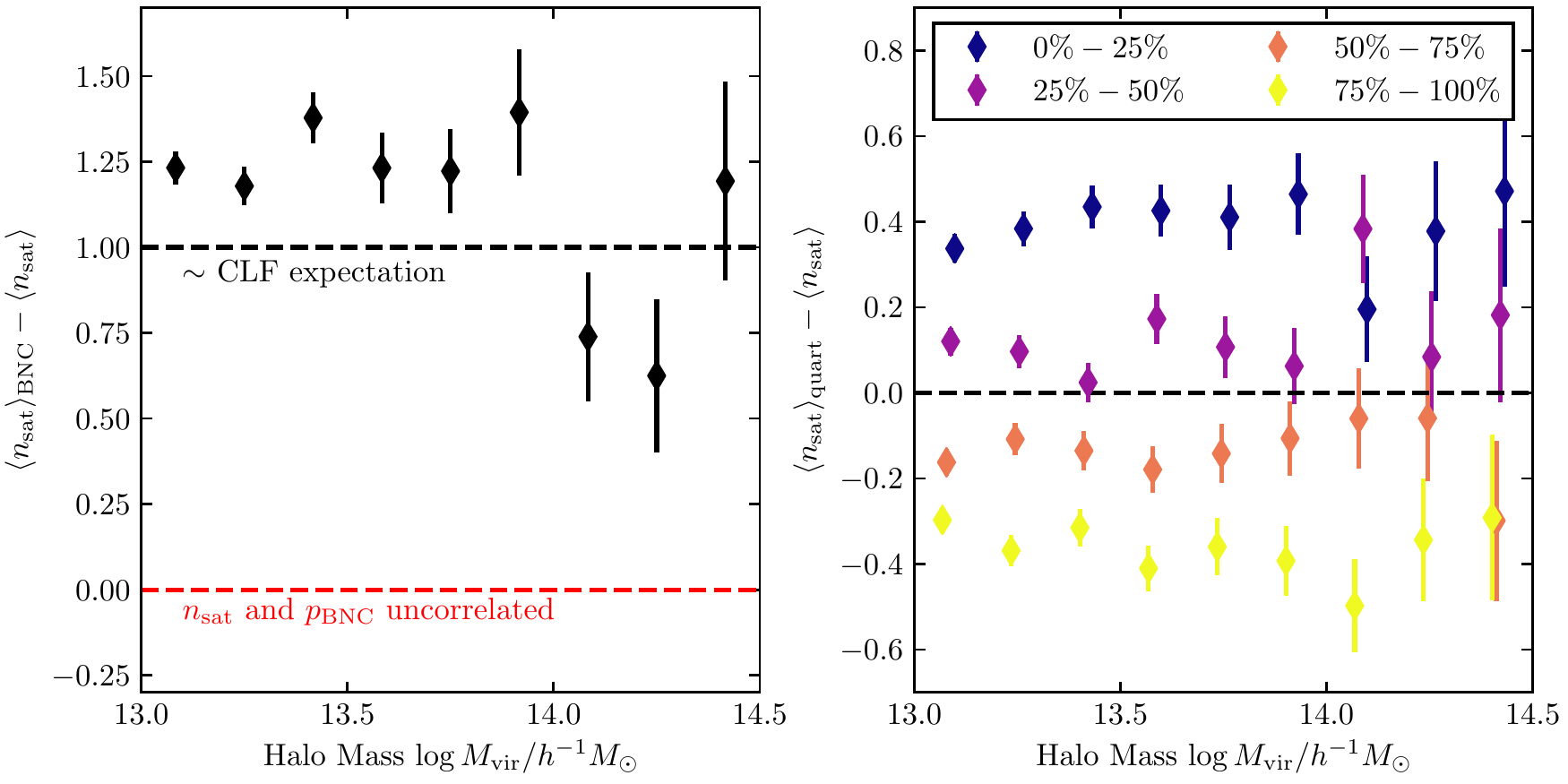} \caption{Results
  from the semi-analytic model of galaxy formation \citep{DeLucia+07}
  run on the Millennium simulation. The left-hand panel shows the
  difference in satellite occupation between haloes where the
  brightest galaxy is a satellite and all haloes. Generally, the
  difference is roughly unity, following the CLF expectation if
  satellite luminosities are uncorrelated and the satellite number
  follows a Poisson distribution. The right-hand panel presents a
  similar occupation difference but now for haloes binned in quartiles
  of their central luminosity. Note that, at fixed halo mass, haloes with a 
  brighter central have fewer satellites.}  \label{fig:SAM}
\end{figure*}

Furthermore, there is a third, less intuitive dependence that is
important to consider. In principle, the probability of a satellite
being brighter than the central should have a correlation with the
satellite occupation. If a halo has more satellites, it should be more
likely that one of those satellites turns out to be brighter than the
central. For example, CLF models commonly assume that satellite
luminosities have no correlations among each other. Furthermore, it is
assumed that the actual number of satellites $n_{\rm sat}$ is related
to the expected number of satellites $\langle n_{\rm sat} \rangle$
above the detection threshold $L_{\rm th}$,
\begin{equation}
\langle n_{\rm sat} \rangle (M) = \int\limits_{L_{\rm th}}^\infty \Phi_\rms (L 
| M) dL,
\end{equation}
via a Poisson distribution. As shown in the appendix, the number $n_{\rm BTC}$
of satellites brighter than the central of luminosity $L_\rmc$ also
follows a Poisson distribution with an expected value
\begin{equation}
\langle n_{\rm BTC} \rangle (L_\rmc, M) = \int\limits_{L_\rmc}^\infty 
\Phi_\rms(L | M) dL.
\end{equation}
Most importantly the number of satellites fainter than $L_\rmc$, for which the 
expectation value is $\langle n_{\rm FTC} \rangle = \int_{L_{\rm th}}^{L\rmc} \Phi_\rms (L | M) dL$, is 
independent of $n_{\rm BTC}$. Hence, we have that $\langle n_{\rm sat} \rangle = 
\langle n_{\rm BTC} \rangle + \langle n_{\rm FTC} \rangle$. If $\langle n_{\rm BTC} \rangle$ is small such 
that $n_{\rm BTC} = 0$ or $1$, we thus see that $n_{\rm sat}$ will on average be higher by $1$ 
for systems where the brightest galaxy is not the central ($n_{\rm BTC} = 1$). 
In other words, the common assumptions of CLF modelling imply a positive
correlation between the probability $p_{\rm BNC}$ that the BHG is the
satellite and the satellite occupation $n_{\rm sat}$. Taking this
correlation between $n_{\rm sat}$ and $p_{\rm BNC}$ into account is
important because we applied a cut on the number of secondaries,
i.e. satellites, when constructing our sample. Given how strongly the
number of primaries decreases for a given cut in the number of secondaries, as 
shown in Figure \ref{fig:Primaries}, our selection is biased towards systems 
where $n_{\rm sat}$ is larger than its expectation value.

While a correlation between satellite occupation and $p_{\rm BNC}$
is implied by our assumed CLF, it does not need to be true for the actual 
Universe. We investigate the realism of such a correlation by analysing the 
galaxy catalogue of the semi-analytic model of \cite{DeLucia+07} which was run 
on the dark matter halo catalogue of the Millennium simulation 
\citep{Springel+05}. The motivation behind using a semi-analytic model is 
to explore statistics in a rich mock galaxy sample for which the correct
answer is known. We first bin haloes in the halo catalogue by their 
host halo mass. For each small bin in host halo mass we determine a luminosity 
threshold such that the average satellite occupation in that halo mass bin is 
$3$. We apply this cut primarily because most primaries in our sample
have roughly $3$ secondaries. We now have a sample of haloes in
which the average number of satellites $\langle n_{\rm sat} \rangle$
is $3$, regardless of halo mass. In the left panel of
Figure \ref{fig:SAM} we show $\langle n_{\rm sat} \rangle_{\rm BNC}
- \langle n_{\rm sat} \rangle$, the difference in satellite occupation
between haloes in which the central is not the BHG and the general
population. If $p_{\rm BNC}$ were uncorrelated with $n_{\rm sat}$,
this difference should be $0$. On the other hand, assuming that
satellite luminosities are uncorrelated and that $\langle n_{\rm
BTC} \rangle \ll 1$ would imply a difference of $\sim 1$. Indeed, we
do find $\langle n_{\rm sat} \rangle_{\rm BNC} - \langle n_{\rm
sat} \rangle \approx 1$ for all halo masses.

Given that central galaxies can grow by the accretion of
satellites, we expect a negative correlation between central
luminosity $L_\rmc$ and satellite occupation $n_{\rm sat}$. To show
this we divide the \cite{DeLucia+07} galaxy catalogue at each halo mass into 
quartiles of $L_\rmc$. The right-hand panel of Figure \ref{fig:SAM} displays the
satellite occupation difference between central galaxies in the
specific quartile compared to all centrals. We find that the most
luminous central galaxies ($75\% - 100\%$) have on average fewer
satellites. Conversely, haloes occupied by the least luminous centrals
($0\% - 25\%$) have a higher satellite occupation on
average. Furthermore, we do expect that $p_{\rm BNC}$ has a
negative correlation with $L_\rmc$, in that it is less likely for
satellites to be brighter than the central if the latter is more
luminous. These anti-correlations between $L_\rmc$ and
$n_{\rm sat}$ and between $L_\rmc$ and $p_{\rm
BNC}$ also imply a positive correlation between $n_{\rm sat}$ and
$p_{\rm BNC}$. However, we have verified that the qualitative trend in
the left-hand panel of Figure \ref{fig:SAM} is not caused by this
correlation with $L_\rmc$. Instead, even when first selecting centrals
in a narrow range of $L_\rmc$, the difference of order unity
remains. We conclude that the positive correlation between $p_{\rm
BNC}$ and $n_{\rm sat}$ implied by CLF models seems to be a good
approximation even for realistic galaxy formation models.

With the above discussion in mind, what is the relation between
$n_{\rm sat}$ and $p_{\rm BNC}$ in CLF models? Once the number of
satellites is specified, the probability that at least one of them is
brighter than the central is given by
\begin{equation}
p_{\rm BNC}(n_{\rm sat} | \langle n_{\rm sat} \rangle, \langle n_{\rm BTC} 
\rangle) = 1 - \left( 1 - \frac{\langle n_{\rm BTC} \rangle}{\langle n_{\rm 
sat} \rangle} \right)^{n_{\rm sat}}.
\label{eq:nsat_Dependence}
\end{equation}
On the other hand, averaged over all realizations of $n_{\rm sat}$ one finds
\begin{align}
f_{\rm BNC} (L_c, M) =& \sum\limits_{n_{\rm sat}} p_{\rm BNC} (n_{\rm 
sat} | \langle n_{\rm sat} \rangle, \langle n_{\rm BTC} \rangle) \, p(n_{\rm 
sat} | \langle n_{\rm sat} \rangle)\nonumber\\
=& 1 - \exp \left[ - \langle n_{\rm BTC} \rangle (L_c, M) \right].
\end{align}
Thus, to model $f_{\rm BNC}$ we need a parametrization for $\langle
n_{\rm BTC} \rangle$. Note that the above equation gives the probably that, in 
a halo of mass $M$, the BHG is not the central, given a central luminosity 
$L_\rmc$, whereas Eq.~(\ref{eq:f_BNC_CLF}) gives the same probability, given a 
BHG of luminosity $L_{\rm BHG}$. Note that, whereas the latter is an observable quantity, 
the former is not, as it requires perfect {\it a priori} knowledge of which galaxy is 
the actual central. Thus, we choose to parametrize $\langle
n_{\rm BTC} \rangle$ as a function of $L_{\rm BHG}$. Under the assumption that 
the halo mass and BHG luminosity dependencies are independent, we therefore 
adopt the following fitting function for $\langle n_{\rm BTC} \rangle$,
\begin{align}
\langle n_{\rm BTC} \rangle (L_{\rm BHG}, M) &=\nonumber\\
f_0 + a_1 &\log L_{10} + a_2 (\log L_{10})^2 + b \log M_{12},
\label{eq:n_BNC}
\end{align}
where $f_0$, $a_1$, $a_2$ and $b$ are free parameters, $M_{12}$
is the halo mass $M_{200b}$ in units of $10^{12} \Msunh$, and $L_{10}$
is the BHG luminosity in units of $10^{10} \Lsunh$. The above parametrization 
is motivated by the $f_{\rm BNC} (L_{\rm BHG}, M)$ dependence for CLF models, 
as described in more detail in section \ref{subsec:CLF_Implications}. Note that
due to our parametrization it is possible that $\langle n_{\rm
BTC} \rangle > \langle n_{\rm sat} \rangle$. In this case we set
$p_{\rm BNC}$ to unity. When constructing mock
catalogues, we compute $p_{\rm BNC}$ for each system via 
Eq.~(\ref{eq:nsat_Dependence}), taking into account both the expected and
actual satellite occupation. We then draw a random number between $0$
and $1$ and if that number is below $p_{\rm BNC}$, we swap the
phase-space coordinates of the central, which was the BHG so far, and
the brightest satellite. Note that although $n_{\rm BTC}$ can be larger than 
unity, this method of swapping central and satellite never sets $n_{\rm BTC}$ 
larger than unity, i.e. the central is at least the second brightest galaxy. 
However, this is not crucial because for our observations we are only concerned 
with the phase-space position of the brightest galaxy and ignore the 
luminosities of secondaries. Finally, note that swapping phase-space positions 
in the above described way does conserve the total CLF, $\Phi_\rmc + 
\Phi_\rms$, and BHG luminosity distribution $\Phi_{\rm BHG}$ for each halo, 
while modifying the central and satellite CLF.

\subsection{Mock SDSS Surveys}

To create mock SDSS-like galaxy surveys we first populate all dark
matter haloes in the SMDPL simulation assuming a constant luminosity
threshold of $10^9 \Lsunh$. We then place a virtual observer
with random orientation at a random position inside the simulation
box. By periodically repeating the galaxy catalogue if needed we
calculate all the galaxies within a comoving distance corresponding to
$z_{\rm max} = 0.17$. For each galaxy we compute its apparent
magnitude, taking into account $k$- and evolution corrections as a function of
redshift. The evolution correction has an analytical form, whereas we
estimate the $k$-corrections by fitting a linear relation to the
$k$-correction of observed galaxies in SDSS \citep{Blanton+07}. We
then require apparent magnitudes $m_r < 17.6$, similar to the SDSS DR7
sample, and apply the SDSS DR7 survey mask. Additionally, we apply
redshift space distortions according to the line-of-sight velocity of
each galaxy. Furthermore, we add a random velocity drawn from a Gaussian with a 
scatter of $35 \, \kms$ to that line-of-sight velocity to account for 
spectroscopic redshift errors in the SDSS \citep{More+09b}. Finally, we 
simulate 
fibre collisions. If two galaxies
are separated by less than $55''$ we assign one of them the redshift
of the nearest neighbour with a probability of $65\%$. This
probability is not unity because some galaxy pairs are observed with
several spectroscopic plates in which case redshifts can be measured
for both despite the small separation.

\subsection{Parameter Inference}

\begin{figure}
  \centering \includegraphics[width=\columnwidth]{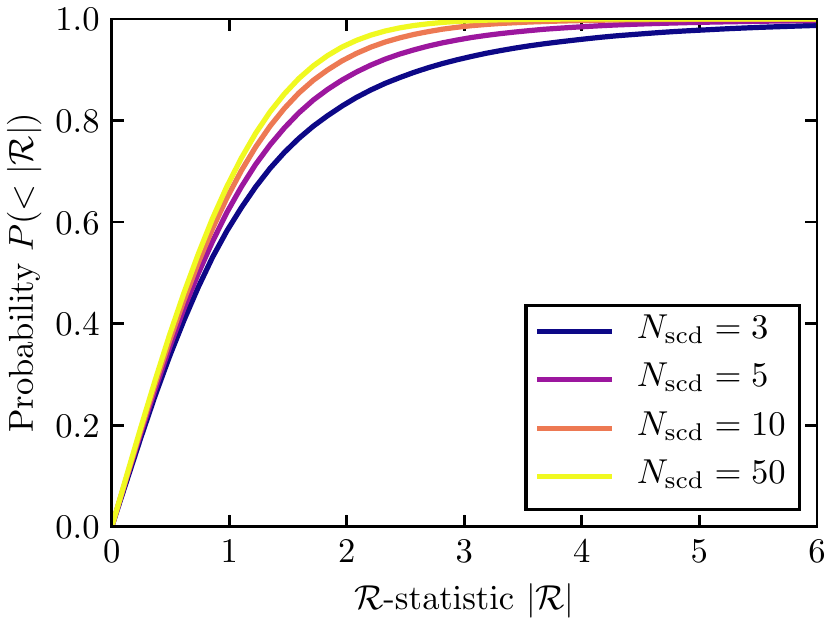} \caption{The
  dependence of the $\mathcal{R}$-statistic on the number of galaxies
  used to compute it. For simplicity, we assume here that the BHG is always the 
  central, i.e. $f_{\rm BNC} = 0$. The line-of-sight velocities of satellites
  are drawn from a single Gaussian. Note that the distribution of the 
  $\mathcal{R}$-statistic has a slight dependence on $N_{\rm scd}$ at fixed 
  $f_{\rm BNC}$.}  \label{fig:R_vs_Nsat}
\end{figure}

We create $25$ mock SDSS-like galaxy catalogues using the procedure
described above. For a given choice of parameters, $\{f_0, a_1, a_2,
b\}$, we can compute the $\mathcal{R}$ and $\mathcal{S}$-statistics in
the mock catalogues. We divide our sample of primaries into a low, $z <
0.09$, and a high, $z \geq 0.09$, redshift sample and $4$ luminosity
bins defined by $10.0, 10.25, 10.5, 10.75$ and $11.25$ in $\log [L_{\rm
BHG} / (\Lsunh)]$. Altogether, we have $8$ sub-samples for which
we compare the distributions of $\mathcal{R}$ and $\mathcal{S}$. For
each of the $8$ sub-samples the distributions of the $\mathcal{R}$ and
$\mathcal{S}$-statistic are derived by merging the distributions of
all $25$ mock catalogues. Note that the $\mathcal{R}$ and
$\mathcal{S}$-statistics at fixed $f_{\rm BNC}$ have a small dependence on the 
number of secondaries used. 
The dependence is such that a lower value of $N_{\rm
scd}$ has an effect similar to a higher value of $f_{\rm BNC}$. This
is demonstrated in Figure \ref{fig:R_vs_Nsat} where we show the
distribution of $\mathcal{R}$ values for different values of $N_{\rm
scd}$. For this plot, the velocities of secondaries were drawn
from a single Gaussian, i.e. we assumed $f_{\rm BNC} = 0$. This introduces the 
possibility that any
miss-match in the $N_{\rm scd}$ distributions between observations and
our mock catalogues could bias our estimate of $f_{\rm
BNC}$.\footnote{We have tested that using the biweight
estimator \citep{Beers+90} for $\hat{\sigma}_{\rm scd}$ does not
improve the situation.} To alleviate the problem we weight each
primary in the mock catalogues when calculating the $\mathcal{R}$ and
$\mathcal{S}$ distributions. The weight for each primary is $N_{i, \rm
obs} / N_{i, \rm mock}$ where $N_{i, \rm obs}$ is the number of
primaries with $i$ secondaries in the SDSS observations and
$N_{i, \rm mock}$ the same quantity for the mock surveys. We do this
weighting in each of the 8 sub-samples separately. We note, however,
that applying this weighting has only a very small influence on the
resulting $\mathcal{R}$ and $\mathcal{S}$-distributions because
$\langle N_{\rm scd} \rangle$ is overall well re-produced in the
mocks.

To quantitatively compare the mock catalogues to the observations we
fit binned distributions for both $|\mathcal{R}|$ and
$|\mathcal{S}|$ in each of the $8$ sub-samples. The bin edges are defined by 
$0$, $0.75$, $1.5$, $3.0$ and $\infty$. Under the assumption that all 
$\mathcal{R}$ and $\mathcal{S}$ values are independent, the probability mass 
function for a single sub-sample is given by a multinomial distribution such 
that the likelihood $\mathcal{L}$ is proportional to
\begin{equation}
\mathcal{L} \propto \prod\limits_{i = 1}^4 P_{i, \rm mock}^{k_{i, \rm obs}},
\end{equation}
where $P_{i, \rm mock}$ is the probability that $|\mathcal{R}|$ or
$|\mathcal{S}|$ are in a given bin as determined from the mock catalogues and 
$k_{i, \rm obs}$ are the corresponding observed numbers.

We use \texttt{MultiNest} \citep{Feroz+08} to
evaluate the parameter uncertainties. We assume a flat prior in $[-3,
3]$ for $f_0$. On the other hand, $a_1$, $a_2$ and $b$ are generally
not perfectly constrained and very large absolute values for these
parameters are permitted by the data. We thus choose flat priors in
$[-1, 1]$ for $\hat{a}_1$, $\hat{a}_2$ and $\hat{b}$ defined by $a_1
= \tan (\hat{a}_1 \pi / 2)$, $a_2 = \tan (\hat{a}_2 \pi / 2)$ and $b
= \tan (\hat{b} \pi / 2)$. This effectively maps $[-1, 1]$ to
$[-\infty, \infty]$. We use $1000$ live points, a target sampling
efficiency of $25\%$ and $\Delta \ln \mathcal{Z} < 0.1$ as a stopping
criterion in all cases. Here, $\mathcal{Z}$ is the estimate for the global 
evidence \citep{Feroz+08}. We have verified that our results are
converged by running \texttt{MultiNest} with $2000$ live points and using $50$ 
mock catalogues and getting equivalent results. Altogether, roughly $60,000$ 
likelihood evaluations are performed to arrive at the posterior distribution.

\section{Results}
\label{sec:Results}

In this section we present our results from the analysis procedure
described in the previous section.

\subsection{Default Analysis}

\begin{figure*}
  \centering \includegraphics[width=\textwidth]{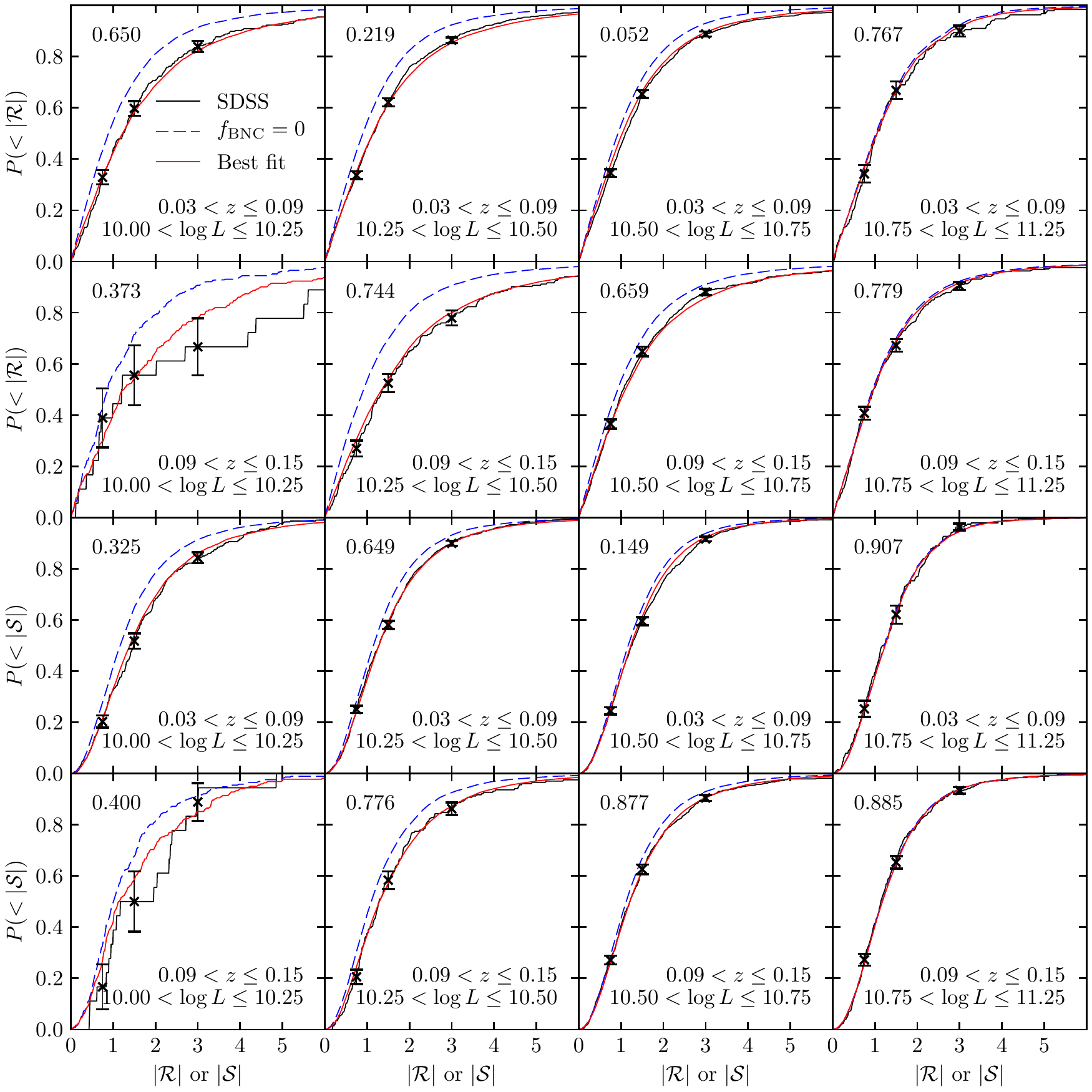} \caption{The
  $\mathcal{R}$ and $\mathcal{S}$ distributions for our mock
  catalogues with $f_{\rm BNC} = 0$ (blue, dashed), the best-fitting
  model (red) and the SDSS data (black with error bars). The upper and 
  lower two rows show the $\mathcal{R}$- and $\mathcal{S}$-statistics, 
  respectively. The luminosity increases from left to
  right. In the lower right corner of each plot we show the redshift range
  and BHG luminosity range in $\log [L_{\rm BHG} /(\Lsunh)]$. 
  Generally, the difference between the $f_{\rm BNC} = 0$
  mocks and the SDSS data decreases with luminosity. The best-fitting
  model can explain the data distribution in all cases. In the upper
  left side of each plot we show the $p$-value returned from a KS test
  comparing the best-fitting distributions of $|\mathcal{R}|$ or
  $|\mathcal{S}|$ against the SDSS measurements. The location of the error bars 
  on the SDSS measurements correspond to the bin edges in $|\mathcal{R}|$ and 
  $|\mathcal{S}|$ used for fitting the data.}  \label{fig:R_and_S}
\end{figure*}

Figure \ref{fig:R_and_S} shows the distributions of the $\mathcal{R}$
and $\mathcal{S}$-statistics for the SDSS data (black with error bars),
the mock catalogues with $f_{\rm BNC} = 0$ for all halo masses and BHG
luminosities (blue, dashed) and the best-fit, non-zero $f_{\rm BNC}$
model (red). We show all the $8$ sub-samples divided into different
BHG luminosities (different columns) and redshifts (different rows) 
that we used during the fitting, as indicated. The error
bars on the SDSS data are computed assuming a binomial distribution,
i.e. $\Delta P = \sqrt{P (1 - P) / n}$, where $n$ is the number of
primaries sampling the distributions. The $x$-positions of the error bars are 
the 
same as the bin edges used to fit the data. Note however that we used the 
binned probability distribution function instead of the cumulative distribution 
function to fit the data.

The first thing to note
is that the relative differences between the $f_{\rm BNC} = 0$ model
and the SDSS data decrease with increasing BHG luminosity. This
indicates that $f_{\rm BNC}$ decreases with BHG luminosity for this
particular sample of systems. Note however that this does not
necessarily imply that $f_{\rm BNC}$, averaged over BHG luminosity,
decreases with halo mass despite luminosity and halo mass being
positively correlated. Instead, we will show later that there is a
strong anti-correlation of $f_{\rm BNC}$ with BHG luminosity at fixed
halo mass. This is the main driver of the apparent decrease of $f_{\rm
BNC}$ with the luminosity of the BHG.

The best-fit model is able to accurately fit all $16$ distributions. We can
gauge the quality of the fit by using a Kolmogorov--Smirnov (KS)
test to compare the best-fit $|\mathcal{R}|$ and $|\mathcal{S}|$
distributions to the SDSS data. The $p$-values of the KS-test are shown
in the upper left corner of each plot. These $p$-values have not been
used in the fit and we quote them here primarily to facilitate a
comparison with \cite{vdBosch+04} and \cite{Skibba+11} who also used
KS tests. Most of the fits are excellent with $p$-values above
$15\%$. Alternatively, we can
calculate a $\chi^2$ value by approximating all data points in the cumulative 
distribution function as a multi-variate Gaussian distribution. Under this 
approximation we get $\chi^2 / \mathrm{dof} = 35.1 / (48 - 4)$, again showing a 
good overall fit. Most importantly, our best-fit
model can explain both the $\mathcal{R}$ and
$\mathcal{S}$-distributions simultaneously for all BHG
luminosities.

Our results are also relevant to the relative relaxation of the systems in our 
samples. As discussed in \cite{Skibba+11}, if central galaxies
would be unrelaxed with respect to the dark matter core, a scenario
which we do not model, the effect on the $\mathcal{R}$-statistic would be much
stronger than on the $\mathcal{S}$-statistic. Thus, our analysis
confirms the findings of \cite{Skibba+11} that the detected
phase-space offsets in the $\mathcal{R}$ and $\mathcal{S}$-statistics
are likely due to satellites galaxies being BHGs.

\begin{figure*}
  \centering \includegraphics[width=\textwidth]{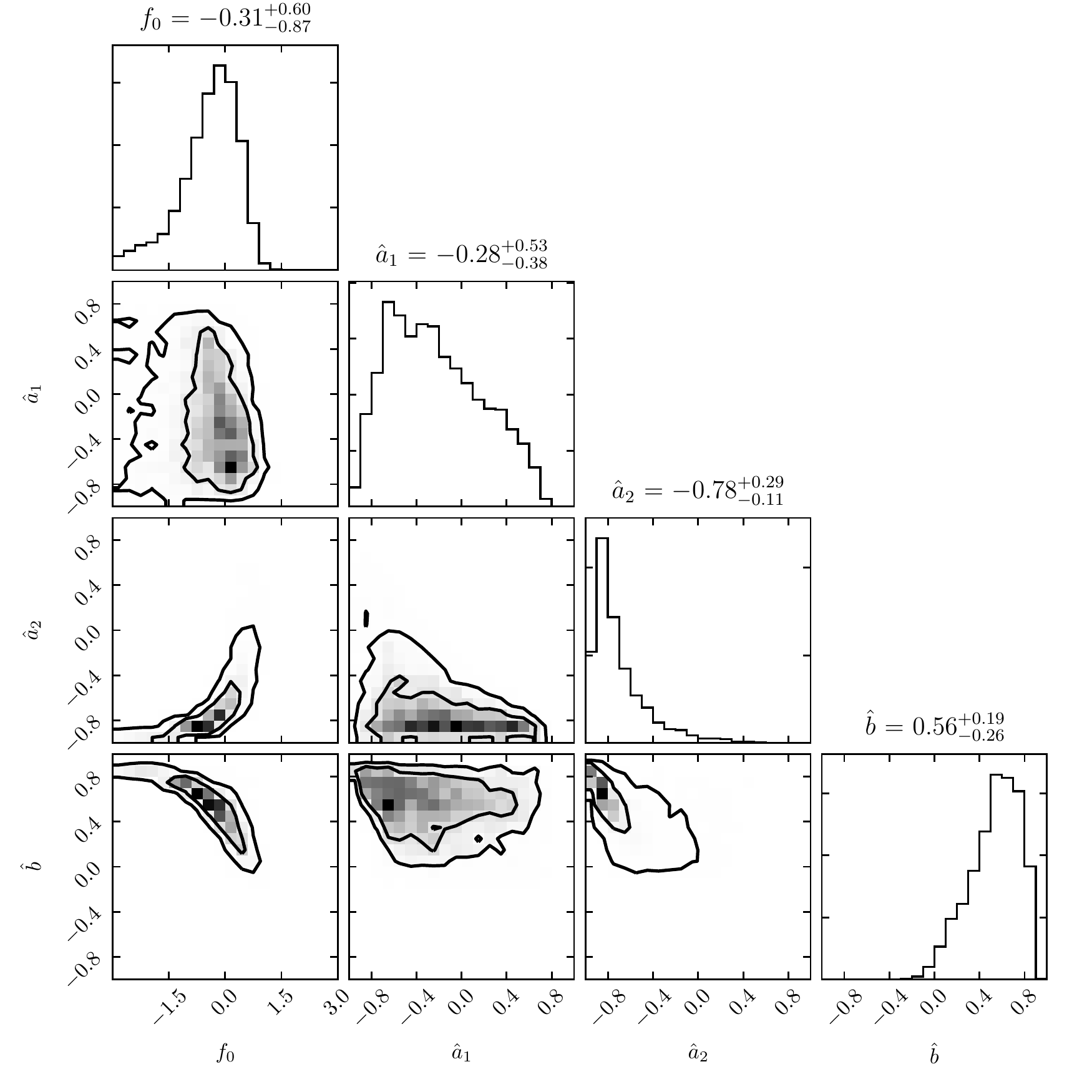} \caption{The
  marginalized posterior distributions of the parameters describing
  $f_{\rm BNC}(L_{\rm BHG}, M)$ for the default analysis. The model has been 
  fitted to the entire SDSS sample. The contour levels on the two-dimensional 
  histograms correspond to the $68\%$ and $95\%$
  confidence intervals. We also show the marginalized one-dimensional
  posteriors for each parameter. Note that we show $\hat{a}_1$,
  $\hat{a}_2$ and $\hat{b}$, where $a_1 = \tan(\hat{a}_1 \pi/ 2)$,
  $a_2 = \tan(\hat{a}_2 \pi/ 2)$ and $b = \tan(\hat{b} \pi /
  2)$.}  \label{fig:corner}
\end{figure*}

The posterior distributions of the parameters are shown in Figure \ref{fig:corner}. The two-dimensional histograms show
distributions for different parameter combinations with $68\%$ and
$95\%$ confidence intervals. Note that we show $\hat{a}_1$,
$\hat{a}_2$ and $\hat{b}$, not $a_1$, $a_2$ and $b$. The default
analysis prefers models in which $n_{\rm BTC}$, and thereby $f_{\rm
BNC}$, decreases with BHG luminosity at fixed halo mass ($a_1 < 0$,
$a_2 < 0$), in agreement with the visual inspection of
Figure \ref{fig:R_and_S}, and increases with halo mass at fixed BHG
luminosity ($b > 0$). Specifically, $b > 0$ is found with $> 98\%$
probability. These constraints on $b$ primarily come from the fact
that we divided our sample into a low and high-redshift sample in
addition to splitting in BHG luminosity. Because the detection
threshold increases with redshift and primaries need to have
at least $3$ detected secondaries, the high-redshift sample probes
systems with higher halo masses on average compared to the
low-redshift sample at the same BHG luminosity.

\begin{figure}
  \centering \includegraphics[width=\columnwidth]{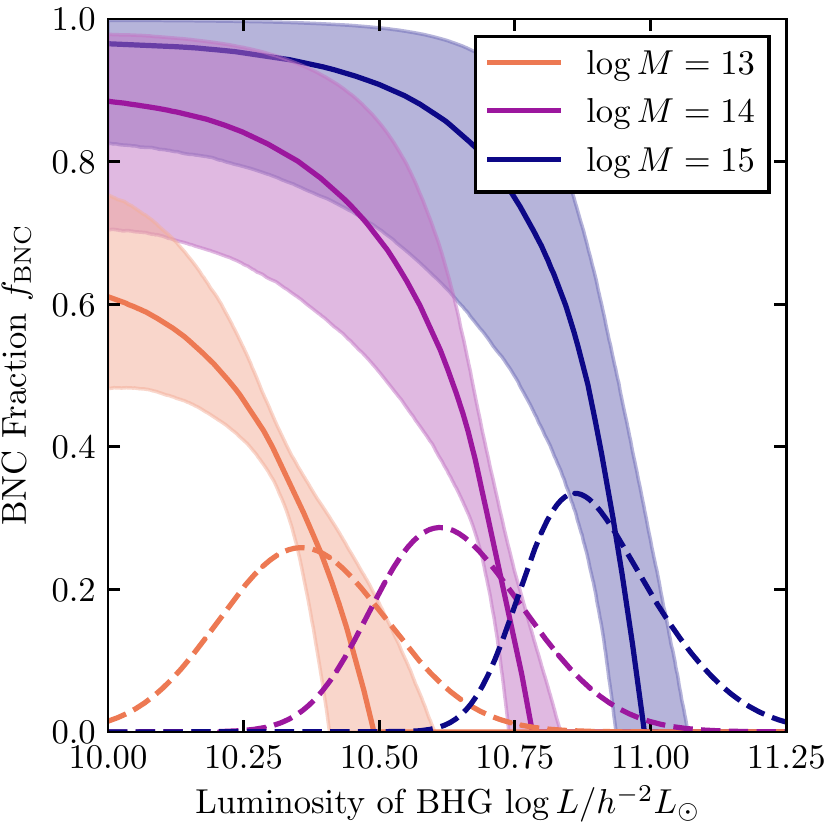} \caption{The
  posterior predictions for the fractions of haloes as a function of
  BHG luminosity where the brightest galaxy is not the
  central. Different colours indicate different halo masses in units of
  $\log [M_{200b} / (\Msunh)]$. Bands are $68\%$ confidence
  intervals. The dashed lines show the distributions of BHG
  luminosities that result from our choice of the CLF
  model.}  \label{fig:BNC_vs_L_1}
\end{figure}

The main result of our work is summarized in
Figure \ref{fig:BNC_vs_L_1}. We show $f_{\rm BNC}$ as a function of
BHG luminosity for different halo masses. The bands indicate $68\%$
confidence intervals and the dashed lines the distributions of BHG
luminosities in our CLF models. Three different trends are apparent
from this figure. Firstly, for a given halo mass the probability that
the brightest galaxy is not the central strongly decreases with
increasing BHG luminosity. This trend is expected because for a low-luminosity 
central galaxy it is more likely that its luminosity is exceeded by that of a satellite 
galaxy. Secondly, for a fixed BHG luminosity $f_{\rm BNC}$ increases with halo 
mass. This trend arises naturally from the fact that the satellite occupation 
or cluster richness increases with halo mass, making it more likely that a 
satellite is brighter than the central if the central luminosity is kept fixed. 
As we will show later, both qualitative results are also predicted by the CLF 
model of \cite{Cacciato+09, Cacciato+13}. Finally, there is a trend of
increasing $f_{\rm BNC}$ with halo mass in general. In particular,
$f_{\rm BNC}$ for the median BHG luminosity increases with halo
mass. We will discuss the trend of $f_{\rm BNC}$ with halo mass in
more detail in section \ref{subsec:Comparison}.

\subsection{Dependence on CLF Parameters}
\label{subsec:CLF_Dependence}

\begin{figure}
	\centering 
	\includegraphics[width=\columnwidth]{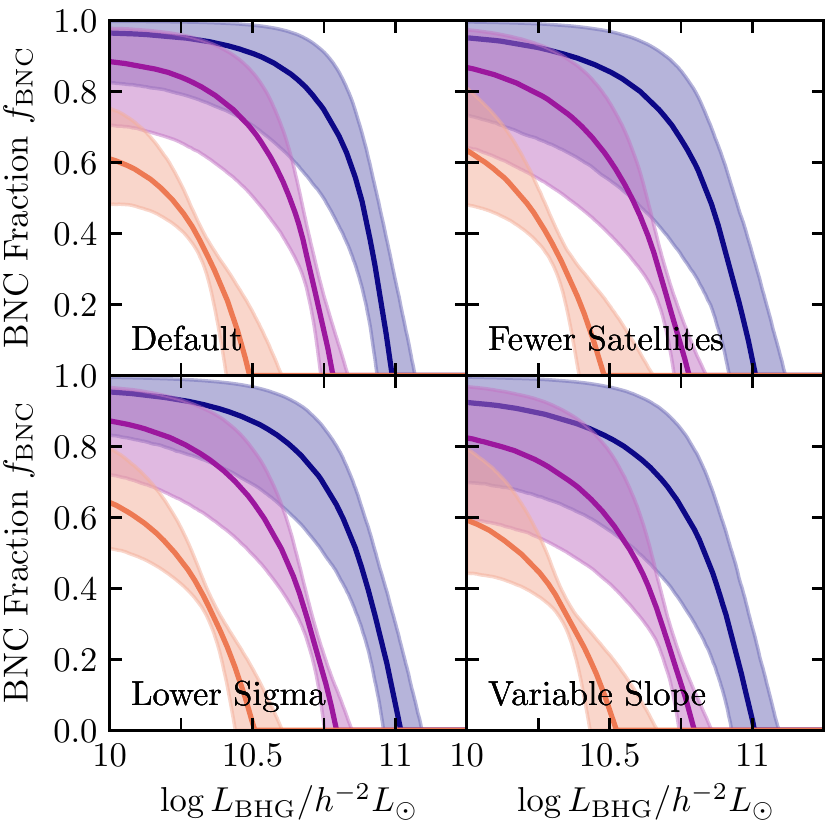} 
	\caption{Similar to Figure \ref{fig:BNC_vs_L_1}, we show the posterior 
	prediction for $f_{\rm BNC}$ as a function of BHG luminosity and halo mass. 
	The colours have the same meaning as in Figure \ref{fig:BNC_vs_L_1}. 
	Different panels are for different CLF choices to create the mock 
	catalogues. All models have been fit to the entire SDSS sample.}  
	\label{fig:BNC_vs_L_Comparison}
\end{figure}

Our results so far have been derived using mock catalogues constructed
with a particular CLF parametrization. Specifically, we used the CLF
parametrization and best-fit parameters
of \cite{Cacciato+13}. However, this model is an imperfect representation of 
the real Universe for several reasons. For example, the best-fit parameters in 
\cite{Cacciato+13} are not tuned to reproduce the observations that we 
consider, such as those of Figure \ref{fig:Primaries}. Also, \cite{Cacciato+13} 
fitted CLF and cosmological parameters simultaneously and the best-fit
cosmological parameters are somewhat different from those adopted for the
SMDPL simulations used to construct our mocks (see \S\ref{sec:Mock
Catalogues}). Further, \cite{Cacciato+13} assumed a different,
more concentrated radial profile for satellites to fit the clustering
of galaxies in SDSS DR7. Finally, there are, of course, statistical and also 
some systematic uncertainties in the CLF parameters derived by 
\cite{Cacciato+13}.

To test the impact of these systematic and statistical uncertainties on our 
results, we run our analysis procedure using mock catalogues with reasonable
variations in the CLF parameters. These different variations are
summarized in Table \ref{tab:CLF_parameters}. The first variation
assumes a lower scatter in the central luminosity at a fixed halo
mass. For this we leave all parameters unchanged but change the value
for $\sigma_{\log L}$ from $0.157 \,{\rm dex}$ to $0.120 \,{\rm dex}$. This value is
roughly $5\sigma$ away from the best-fit value found
in \cite{Cacciato+13}, but systematically lower values for $\sigma_{\log L}$ were 
inferred in analyses which omitted the Sloan Great Wall. In this
case, the authors found $\sigma_{\log L} = 0.142_{-0.009}^{+0.010}$. Thus, a
value as low as $0.120$ is plausible. The second variation
has to do with the slope of the satellite CLF. Generally, the
satellite CLF is given by Eq.~(\ref{eq:CLF_sats}), where
$\alpha_\rms$ is a free parameter in the analysis
of \cite{Cacciato+13}. This implies that the low-luminosity slope of
the satellite CLF is independent of halo mass
$M$. However, in an earlier paper \cite{Cacciato+09} introduced a 
parametrization with a variable luminosity slope,
\begin{equation}
\alpha_\rms (M) = -2.0 + a_1 \left( 1 - \frac{2}{\pi} \arctan \left[ a_2 \log (M / M_2) \right] \right),
\end{equation}
where $a_1$, $a_2$ and $M_2$ are free parameters. A mass-independent slope is
realized by setting $a_2 = 0$. In this second variation of the CLF
parameters we adopt the best-fit $a_1$, $a_2$, $M_2$, $b_0$, $b_1$ and
$b_2$ from \cite{Cacciato+09} for the WMAP3
cosmology \citep{Spergel+07}. Note that \cite{Cacciato+09} used a
different halo mass definition, but we do not apply a correction
because the differences are small, typically $\sim 0.01 \ \mathrm{dex}$. The 
last CLF parametrization assumes a lower overall number of satellites. To do 
this we change $b_0$ from $-1.17$ to $-1.37$ while leaving all other parameters
unchanged. This is motivated by the finding that our mock catalogues
produce more primaries with a given number of secondaries than in the SDSS.

\begin{table*}
\begin{tabular}{c | c c c c c c | c c c c}
Name & \multicolumn{6}{|c|}{CLF Parameters} & \multicolumn{4}{|c|}{$f_{\rm BNC}$ Posteriors} \\
& $\sigma$ & $a_1$ & $a_2$ & $b_0$ & $b_1$ & $b_2$ & $f_0$ & $\hat{a}_1$ & $\hat{a}_2$ & $\hat{b}$ \\
\hline\hline
Default & $0.157$ & $0.82$ & $0.00$ & $-1.17$ & $1.53$ & $-0.217$ & 
$-0.31_{-0.87}^{+0.60}$ & $-0.28_{-0.38}^{+0.53}$ & $-0.78_{-0.11}^{+0.29}$ & 
$0.56_{-0.26}^{+0.19}$ \\
Lower Sigma & $0.120$ & $0.82$ & $0.00$ & $-1.17$ & $1.53$ & $-0.217$ & 
$0.02_{-0.42}^{+0.41}$ & $-0.51_{-0.26}^{+0.45}$ & $-0.67_{-0.15}^{+0.38}$ & 
$0.50_{-0.21}^{+0.18}$ \\
Variable Slope & $0.157$ & $0.50$ & $2.11$ & $-0.77$ & $1.01$ & $-0.094$ & 
$0.07_{-0.44}^{+0.33}$ & $-0.44_{-0.33}^{+0.48}$ & $-0.62_{-0.19}^{+0.32}$ & 
$0.45_{-0.26}^{+0.24}$ \\
Fewer Satellites & $0.157$ & $0.82$ & $0.00$ & $-1.37$ & $1.53$ & $-0.217$ & 
$0.10_{-0.57}^{+0.45}$ & $-0.56_{-0.26}^{+0.50}$ & $-0.54_{-0.26}^{+0.52}$ & 
$0.50_{-0.32}^{+0.21}$ \\
\hline
\end{tabular}
\caption{Parameters of the different CLF variations and their resulting 
marginalized posteriors for the parameters describing $f_{\rm BNC}$. All models 
have been fit to the entire SDSS sample. The `Default' case represents the main 
result of our analysis.}
\label{tab:CLF_parameters}
\end{table*}

We produce mock catalogues for all three CLF parametrizations and run
the analysis in exactly the same way as for the default analysis. The
marginalized posteriors for the parameters describing $f_{\rm BNC}$
are also listed in Table \ref{tab:CLF_parameters}. Additionally, we show the 
posterior prediction for $f_{\rm BNC}$ as a function of halo mass and BHG 
luminosity for all CLF variations in Figure \ref{fig:BNC_vs_L_Comparison}. We 
do not find that any of our qualitative results are impacted by alternative CLF
choices. While some of the posteriors change slightly for different
CLF models, none of the quantitative results change
significantly. Only for the \texttt{Fewer Satellites} model we find
that the best-fit $f_{\rm BNC}$ as a function of halo mass is roughly
$5\%$ lower. This is likely caused by this model producing a larger
number of interlopers which have a similar effect as increasing
$f_{\rm BNC}$.

\section{Discussion}
\label{sec:Discussion}

\subsection{Comparison to Previous Work}
\label{subsec:Comparison}

\begin{figure}
  \centering \includegraphics[width=\columnwidth]{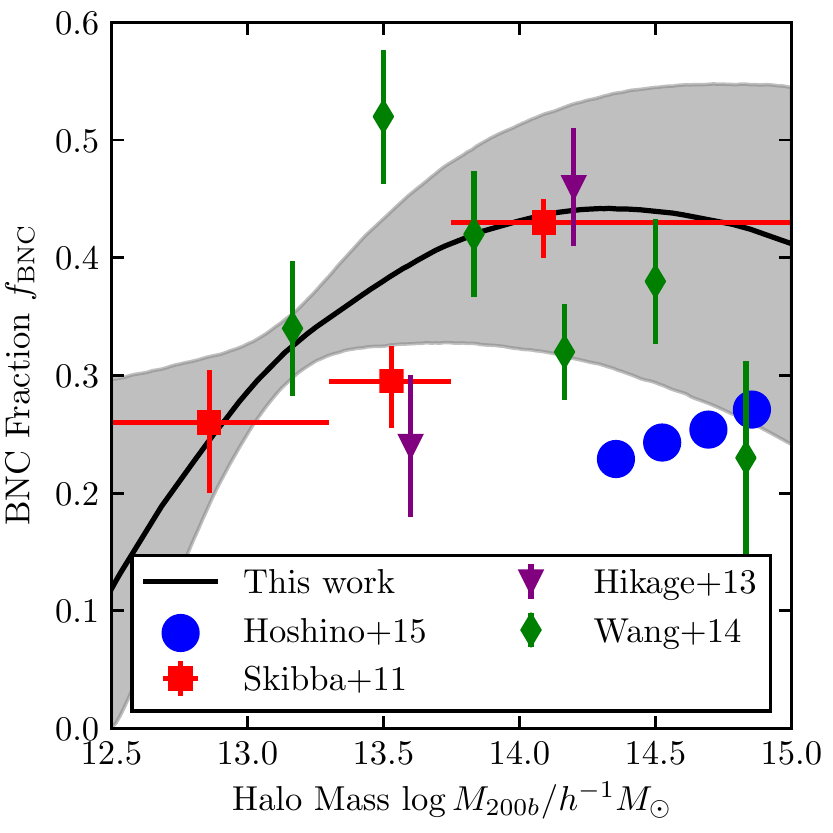} \caption{The
  fraction of systems in which the BHG is not the central as a
  function of halo mass. We compare our results (black, $68\%$
  confidence) to the findings of \protect\cite{Skibba+11} (red squares), 
  \protect\cite{Hikage+13} (purple triangles), 
  \protect\cite{Wang+14} (green diamonds) and \protect\cite{Hoshino+15} (blue 
  dots). Note that \protect\cite{Wang+14} determined how often the most massive 
  galaxy is not the central whereas all other data points are for the most 
  luminous. Error bars were added to the results of \protect\cite{Wang+14} 
  assuming a binomial distribution, i.e. $\Delta f_{\rm BNC} = \sqrt{f_{\rm 
  BNC} (1 - f_{\rm BNC}) / N}$, where $N$ is the number of groups analysed. The 
  statistical uncertainties in the findings of \protect\cite{Hoshino+15} were 
  negligible and are not included in this figure. See the text for a detailed 
  discussion of these results in the context of our findings.}  
  \label{fig:BNC_vs_Mvir}
\end{figure}

Many previous studies have analysed the dynamical state of the
BHGs \citep{vdBosch+05, Sanderson+09, Zitrin+12}. In
particular, \cite{Skibba+11}, \cite{Hikage+13}, \cite{Wang+14} and 
\cite{Hoshino+15} have inferred
$f_{\rm BNC}$ as a function of halo mass. To compare our result to
these studies we marginalize the results shown in
Figure \ref{fig:BNC_vs_L_1} over the BHG luminosity distribution for
each halo mass. The result is shown in
Figure \ref{fig:BNC_vs_Mvir}. Note that all studies use halo mass definitions 
slightly different from $M_{200b}$, but we have not applied a correction 
because the differences are sufficiently small.

The analysis of \cite{Skibba+11} is very similar to our work. In
particular, \cite{Skibba+11} used mock galaxy catalogues to fit the
$\mathcal{R}$ and $\mathcal{S}$ distributions. The main difference is
in their use of the \cite{Yang+07} group finder to identify galaxies
belonging to the same halo and to measure halo
masses. Additionally, \cite{Skibba+11} did not characterize the
dependence of $f_{\rm BNC}$ on BHG luminosity, and did not model
several effects that we consider in this work, like the velocity
offsets of halo cores, velocity correlations between satellites or the
correlation of $n_{\rm sat}$ and $p_{\rm BNC}$. Nevertheless, our
results generally agree very well with those found by \cite{Skibba+11}
at all halo masses. As discussed in section \ref{sec:Mock Catalogues}, we have improved upon the work of \cite{Skibba+11} by accounting for (i) velocity offsets between haloes and centrals, (ii) the unrelaxed state of the satellite population, (iii) correlation between $p_{\rm BNC}$ and the number
of satellite galaxies, (iv) relaxing the assumption that satellite distributions are spherically symmetric and by (v) using a more conservative sample of primaries and secondaries. While many of these effects have a considerable impact on the $\mathcal{R}$ and $\mathcal{S}$ statistics individually (in many cases, the impact exceeds $1\sigma$ significance), they happen to roughly cancel each other, such that our final constraints are very similar to those by \cite{Skibba+11}. A more detailed analysis of the apparent agreement with \cite{Skibba+11} would involve a study of \cite{Yang+07} group finder errors and is beyond the scope of this work.

$f_{\rm BNC}$ has been inferred for Luminous Red Galaxies (LRGs) from the SDSS 
DR7 \citep{Hikage+13}. The authors used the redshift-space power spectrum, 
LRG-galaxy weak lensing and the cross-correlation of LRGs with photometric 
galaxies to infer the fraction of off-centred LRGs. The analysis was done separately for 
systems with one LRG and for systems with multiple LRGs, as determined from a 
Counts-in-Cylinders technique. The strongest constraints for the off-centring 
fraction came from the cross-correlation with photometric galaxies. It was 
found that $24 \pm 6 \%$ of LRGs in single-LRG systems and $46 \pm 5 \%$ of the 
brightest LRGs in multiple-LRG systems are off-centred, i.e. $f_{\rm BNC} = 24 
\pm 6 \%$ and $46 \pm 5 \%$ respectively. These values are shown in Figure 
\ref{fig:BNC_vs_Mvir} together with the average halo masses estimated from 
gravitational lensing \citep{Hikage+13}. Whereas the study by \cite{Hikage+13} 
focuses exclusively on haloes hosting LRGs as defined by the colour and 
magnitude cuts described in \cite{Eisenstein+01}, our results are for all 
galaxies.  Although it is non-trivial to compare results for these different 
selection criteria without a more detailed analysis, we speculate that the main 
impact is as follows; by restricting their analysis to systems that host at 
least one LRG, i.e. a galaxy with an absolute magnitude $M_g < -21.2$, the 
sample analysed by \cite{Hikage+13} is likely biased towards systems with high 
$L_{\rm BHG}$ at fixed halo mass. Our analysis shows that these are systems for 
which $f_{\rm BNC}$ is relatively small. Particularly, we refer the reader to 
section \ref{sec:Satellite_Kinematics} where we investigate $f_{\rm BNC}$ for 
haloes selected purely based on $L_{\rm BHG}$. Hence, we suspect that the 
results of \cite{Hikage+13} most likely represent a lower limit for the 
$f_{\rm BNC}$ of {\it all} galaxies. In light of this, and given the errors on 
their and our measurements, we conclude that the results presented by 
\cite{Hikage+13} are in reasonable agreement with those derived here.

\cite{Wang+14} studied the positions of X-ray peaks in clusters identified by 
the \cite{Yang+07} group finder in SDSS DR7. In their analysis, they denoted 
the central as the galaxy that is closest to the X-ray peak. Additionally, they 
required the central to be among the four most massive (in terms of stellar 
mass) galaxies in the cluster. They find that in roughly $35\%$ of their halos 
the central defined in this way is different from the most massive galaxy. We 
show in Figure \ref{fig:BNC_vs_Mvir} the results for their sample of clusters 
with secure X-ray detections (Sample II) as a function of halo mass as 
determined from the \cite{Yang+07} group catalogue. Note that \cite{Wang+14} 
determined the fraction of haloes in which the most massive galaxy is not the 
central whereas we are concerned with the most luminous. However, in $\sim 
90\%$ of the groups the most luminous galaxy is also the most massive 
\citep{Yang+08, Li+14}. Overall, our findings are in good agreement with those 
found by \cite{Wang+14} using a completely different method. However, also note 
that the position of the X-ray peak does not necessarily always coincide with 
the halo centre \citep{Cui+16}.

The off-centring of BHGs from group centres in the Galaxy And Mass Assembly 
survey  \citep[GAMA, ][]{Driver+11} has been examined by 
\cite{Oliva-Altamirano+14}. The authors find off-centring fractions ranging from 
$10$ to $17\%$ for halo masses of $10^{13}$ to $\sim 4 \times 10^{14}$ using the 
GAMA galaxy group catalogue \citep{Robotham+11}. However, the definition of the 
group centre in the group catalogue is different from the true dark matter halo 
centre. As described in \cite{Robotham+11}, the group centre is determined in 
an iterative fashion. At each step, the centre of light of all group members is 
computed and the most distant group member is rejected in subsequent 
iterations. This is done until two members are left from which the brighter 
group member is determined to be the group centre. We have tested that this 
algorithm is strongly biased towards declaring the BHG the group centre even if 
the BHG is a satellite of the dark matter halo of that group. Thus, the results 
of \cite{Oliva-Altamirano+14} should be regarded as a lower limit on $f_{\rm 
BNC}$ with respect to the dark matter halo centre.

\cite{Hoshino+15} analysed $f_{\rm BNC}$ in the LRG SDSS sample using the 
redMaPPer cluster catalogue. Specifically, the redMaPPer cluster finding 
algorithm determines for each cluster member the probabilities that it is a 
true member (as opposed to an interloper) and that it is the central. 
Taken at face value, these can be used, in a straightforward manner, to
determine for each cluster the probability that its brightest member
is not the central. \cite{Hoshino+15} analysed the halo mass range
$10^{14.5} - 10^{15.5} \Msunh$ and find $f_{\rm BNC} \approx
20\% - 25\%$ for halo masses up to $10^{15.0} \Msunh$. In
the same halo mass range our results suggest $f_{\rm BNC} \approx
40 \pm 15 \%$. Thus, our results are different from the findings
of \cite{Hoshino+15} at the $1.5\sigma$ level. Note that
\cite{Hoshino+15} rely entirely on the cluster membership and central 
probabilities determined by redMaPPer. Recently, \cite{Zu+16} have shown that 
the cluster membership probabilities in redMaPPer can be biased due to
contamination from interloper galaxies, which could potentially impact
the findings of \cite{Hoshino+15}. On the other hand, \cite{Hikage+17}
presented evidence supporting the accuracy of central probabilities in 
redMaPPer. Finally, the results of \cite{Hoshino+15} also depend on the 
richness-based halo mass estimates of redMaPPer. The authors have 
shown that their results for $f_{\rm BNC}$ can range from $\sim 10\%$ to $\sim 40\%$ if an additional dependence of the halo mass estimator on the central 
luminosity is factored in.

It is also possible that part of the discrepancy arises from
the SDSS, as discussed in \cite{Hoshino+15}. The SDSS DR7 pipeline
significantly underestimates the luminosity of luminous extended
galaxies, especially massive elliptical or cD
galaxies \citep{Abazajian+09, Aihara+11, Bernardi+13}. This could lead to an
increased fraction of $f_{\rm BNC}$ observationally. \cite{Hoshino+15}
argue that part of the disagreement between their results (using SDSS
DR8) and \cite{Skibba+11} (using SDSS DR4) might be attributed to this
bias. However, using SDSS DR7 we mostly confirm the findings
of \cite{Skibba+11} and the improvement in the bias going from DR7 to
DR8 is `subtle at best' \citep{Aihara+11}. Finally, while we have tested our results with respect to reasonable changes in the CLF parameters, they might also be affected by our choice of cosmology or the way we place satellites in dark matter haloes. For example, we made the choice of assigning the most luminous satellites the highest $M_{\rm peak}$ subhaloes. Other proxies of satellite luminosity will results in different radial profiles \citep{vdBosch+16} and could also impact our inferences. Such an analysis is beyond the scope of this work.

\subsection{Comparison to SAMs}

\cite{Skibba+11} have shown that their inferred values for $f_{\rm BNC}$
as a function of halo mass are significantly higher than expected from
theoretical models of galaxy formation. In
particular, \cite{Skibba+11} showed that the predictions of the SAMs
of \cite{Croton+06} and \cite{Monaco+07} are roughly a factor of two
lower. Since our values for $f_{\rm BNC}$ are in good agreement
with \cite{Skibba+11}, those conclusions also hold for our
analysis. In addition, we have investigated $f_{\rm BNC} (M_{\rm
vir})$ for the model of \cite{DeLucia+07} and find values decreasing
from $\sim 18\%$ to $\sim 10\%$ when going from a halo mass of
$10^{13}$ to $10^{15} \Msunh$. These values are quite similar to 
those of \cite{Croton+06}. Thus, the \cite{DeLucia+07} model also predicts 
substantially lower values for $f_{\rm BNC}$ than we infer in this
work.

Hence, our findings might reveal an actual shortcoming of galaxy formation
models. \cite{Skibba+11} argue that SAMs often adopt a dynamical
friction time-scale that is too short. That would result in
satellites being disrupted too early, particularly for the more massive 
satellites which should have the shortest dynamical friction timescales. This 
will boost the luminosity (and stellar mass) gap between central and
brightest satellites, and thus reduce $f_{\rm BNC}$. However, note
that \cite{DeLucia+07} use a dynamical friction time-scale that is
twice the classical dynamical friction time-scale \citep{Binney+87}
used by \cite{Croton+06}. Nevertheless, both produce similar
predictions for $f_{\rm BNC}$. Other possibilities for an
under-prediction of $f_{\rm BNC}$ could be an over-quenching of
satellites \citep[e.g.,][]{Weinmann+06} or an incorrect treatment of 
how centrals accrete stellar mass from disrupted satellites. It 
remains to be seen how other models, and hydrodynamical simulations 
of galaxy formation, fare in this respect.

\subsection{Implications for CLF Models}
\label{subsec:CLF_Implications}

\begin{figure}
  \centering \includegraphics[width=\columnwidth]{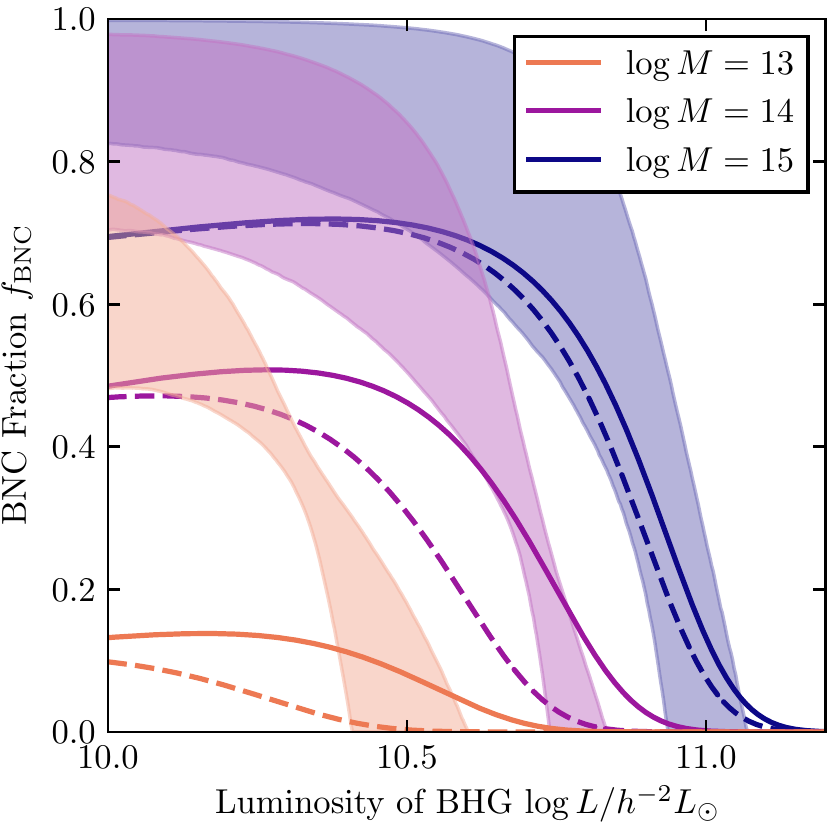} \caption{The
  fraction of systems in which the brightest galaxy is not the central
  as a function of BHG luminosity for different halo masses. Similar
  to Figure \ref{fig:BNC_vs_L_1}, bands show the $68\%$ posterior
  prediction of our analysis using the whole SDSS sample. Dashed lines are the 
  CLF predictions for our default model. Solid lines represent the CLF 
  prediction from an optimized CLF model. See the text for details.}  
  \label{fig:CLF_Prediction}
\end{figure}

For a given CLF model we can compute $f_{\rm BNC}$ as a function of
$L_{\rm BHG}$ and $M_{200b}$ via Eq.~(\ref{eq:f_BNC_CLF}). In
Figure \ref{fig:CLF_Prediction} we show, similar to
Figure \ref{fig:BNC_vs_L_1}, $f_{\rm BNC}$ as a function of $L_{\rm
BHG}$ for different halo masses. Bands shows the $68\%$ posterior
prediction from our analysis and the dashed lines are the predictions
from the default CLF parametrization.

We first note that the CLF model
reproduces the observations on a qualitative level. $f_{\rm BNC}$
increases with halo masses both in the CLF predictions and for our
best-fit model. Also, the luminosity dependence for a fixed halo mass
is quite similar in both cases. However, the CLF model persistently
under-predicts $f_{\rm BNC}$, especially at lower halo masses. This is
similar to the findings of \cite{Skibba+11}. We conclude that our
default CLF model, the one used in \cite{Cacciato+13}, predicts a
$f_{\rm BNC}(M,L_{\rm BHG})$ that is incompatible with the constraints
derived here. In the following, we discuss several possible reasons
for this discrepancy.

The CLF model of \cite{Cacciato+13} assumes that the central
luminosity and the exponential cut-off luminosity of the satellite CLF
have the same ratio for all masses, as given by 
Eq.~(\ref{eq:Cut-off_luminosity}). This ratio has not been fitted in the
analysis of \cite{Cacciato+13}, but has been fixed to the ratio found
in \cite{Yang+08} from an analysis of the \cite{Yang+07} galaxy group
catalogue. In that analysis, though, \cite{Yang+08} assumed that the
most luminous (or most massive) galaxy is always the central and used
this to decompose the CLF into a central and satellite component. In
other words, the authors assumed $f_{\rm BNC} = 0$. Thus, the low
prediction for $f_{\rm BNC}$ from the CLF model might to some extent
be a result of the input assumption of \cite{Yang+08}.

Recently, \cite{Trevisan+17} have analysed the statistics of magnitude gaps in the \cite{Yang+07} SDSS group catalogue. They found that the CLF parametrization of \cite{Yang+08} underpredicts the number of small-gap groups (SGG). They argue that a CLF model that allows for two central galaxies\footnote{We note that in our analysis a group can, by definition, only have one central.} is in much better agreement with observations. While we have not analysed gap statistics, it is plausible that SGGs and groups where the BHG is not the central are closely connected.

It is also possible that some of the other intrinsic assumptions of CLF modelling are
violated. One of those assumptions is that the central luminosity and
the satellite CLF are independent at a fixed halo mass. However, we
have shown that the SAM of \cite{DeLucia+07} predicts an
anti-correlation between satellite occupation and central
luminosity (cf. right-hand panel of Figure \ref{fig:SAM}). Indeed, \cite{Hoshino+15} have argued that their results
for the central occupation of redMaPPer clusters require such an
anti-correlation in order to be reconciled with clustering
studies. Furthermore, the authors state that preliminary studies of
the CLF of redMaPPer clusters provides further evidence for such a
scenario. Any kind of correlation between central luminosity and
satellite occupation would certainly change the $f_{\rm BNC}$
prediction of CLF models. Similarly, a correlation between central and 
satellite luminosities could alter $f_{\rm BNC}$. Another common assumption of 
CLF models is that the satellite numbers follow a Poisson
distribution. However, \cite{Jiang+16} have shown that subhaloes,
which are believed to host satellite galaxies, do {\it not} obey
Poisson statistics \citep[see also][]{Boylan-Kolchin+10, Mao+15}.
Instead, the distribution of the most-massive subhaloes is
sub-Poissonian. If the same is true for the most luminous satellites,
it might also have an impact on the $f_{\rm BNC}$ prediction of CLF
models. It is also worth noting that neither of these effects have
been modelled in our mock catalogues. Thus, they could also alter the
derived values of $f_{\rm BNC}$ of our analysis.

If one wants to keep the traditional assumptions of CLF modelling, we
can try to adjust our CLF parametrization to correctly predict $f_{\rm
BNC}$. Particularly, we explore how changing the constant $0.25$ in Eq. 
(\ref{eq:CLF_sats}) changes the predictions of the CLF. One way would be to 
re-parametrize the satellite CLF via
\begin{equation}
\Phi_\rms (L|M) = \frac{\Phi_\rms^\star(M)}{L_\star} \left( \frac{L}{L_\star} \right)^{\alpha_\rms} \exp\left[ - \delta(M) \left( \frac{L}{L_\star} \right)^2 \right]\,,
\label{eq:CLF_sats_new}
\end{equation}
where we have introduced
\begin{equation}
\log \delta(M) = \delta_1 + \delta_2 \log [M_{200b} / (10^{12} \Msunh)]\,,
\end{equation}
a mass-dependent rescaling of the cut-off scale, and $\delta_1$ and
$\delta_2$ as new free parameters. The original parametrization of \cite{Yang+07}  is recovered for $\delta_1 = \delta_2 = 0$. Experimenting with different values, we find that with $\delta_1 = -0.5$
and $\delta_2 = 0.15$ the CLF predictions are in better agreement with
the posterior predictions, as shown by the solid lines in
Figure \ref{fig:CLF_Prediction} \citep[see also][for analogous conclusions 
obtained with a similar CLF modification to the one shown above]{More+12}. 
However, the CLF model is unable to
reproduce the strong anti-correlation between $f_{\rm BNC}$ and
$L_{\rm BHG}$ for low-mass haloes ($\sim 10^{13} \Msunh$). Even when setting $\delta(M) = 0$, effectively removing the
exponential cut-off, the strong increase of $f_{\rm BNC}$ with
decreasing $L_{\rm BHG}$ at low halo masses is not achieved. In this
case, the aforementioned anti-correlation between central luminosity
and satellite occupation might be able to better explain this result.

To summarize, the best-fit model of \cite{Cacciato+13} underpredicts $f_{\rm BNC}$ as a function of BHG luminosity and halo
mass. It remains to be seen how the CLF model needs to be modified
in order to accommodate the high values of $f_{\rm BNC}(M,L_{\rm BHG})$
inferred here. Most likely, such a modification will affect other
statistics that depend on the CLF, such as galaxy clustering and
galaxy-galaxy lensing. Whether a CLF model can be found that
can simultaneously reproduce all these observables requires a joint
analysis, which is beyond the scope of this work.

\subsection{Satellite Weak Lensing}

\begin{figure}
	\centering 
	\includegraphics[width=\columnwidth]{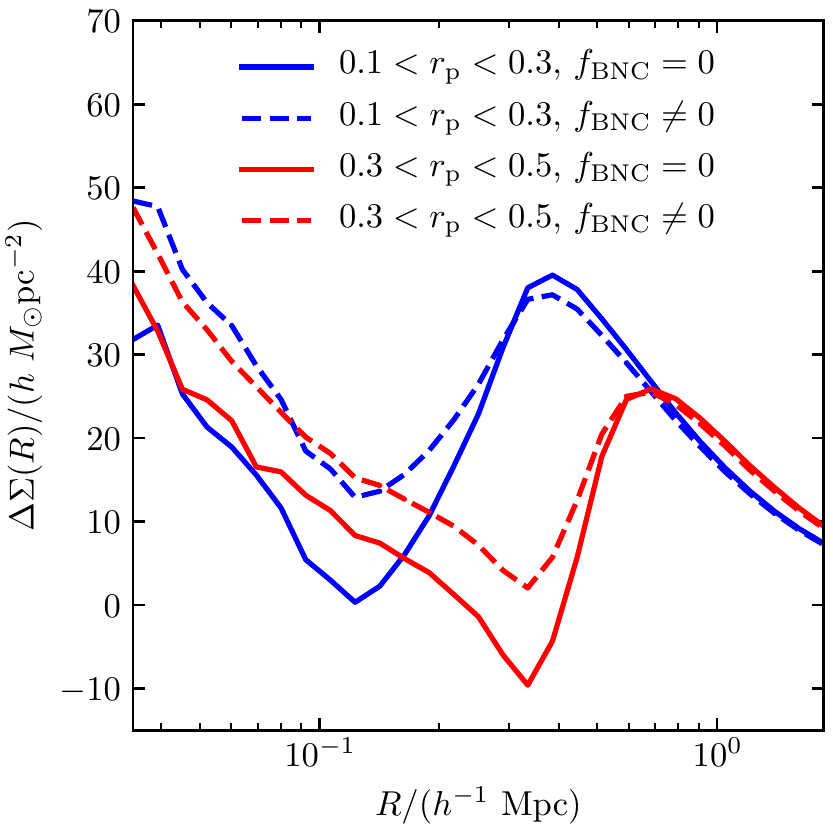}
	 \caption{The excess surface mass density as a function of projected 
	 separation $R$ from the lens galaxy. For this plot we created a mock 
	 galaxy catalogue by populating the halo catalogue of the Bolshoi 
	 simulation according to our default CLF model. We concentrate on galaxies 
	 in host halos of masses $10^{13} - 5 \times 10^{14} \Msunh$. We split 
	 those galaxies by the projected separation to the BHG, $0.1 < r_p / 
	 (\Mpch) < 0.3$ (blue) and $0.3 < r_p / (\Mpch) < 0.5$ (red). Additionally, 
	 we show predictions for a model with 
	 $f_{\rm BNC} = 0$ (solid) and the best-fitting, non-zero model (dashed). 
	 In both cases, a non-zero $f_{\rm BNC}$ leads to an increase of the signal 
	 at small scales.}  
	 \label{fig:Satellite_Weak_Lensing}
\end{figure}

Gravitational lensing has been used successfully to measure the dark 
matter halo masses of satellite galaxies, both in the weak \citep[e.g.,][]{Li+14, Sifon+15}
and strong \citep[e.g.][]{NatarajanKneib97, Natarajan+17} lensing regimes.
Recently, several studies have used weak lensing to infer the stellar-to-halo mass ratio of 
satellite galaxies as a function of projected separation $r_p$ from the halo 
centre \citep{Sifon+15, Li+16, Niemiec+17}, finding evidence of tidal stripping 
near the halo centre.
 
The main observable in the above mentioned studies is the excess surface mass 
density profile $\Delta \Sigma(R)$ around satellite galaxies,
\begin{equation}
\Delta \Sigma(R) = \bar{\Sigma}(<R) - \bar{\Sigma}(R),
\end{equation}
where $R$ is the projected distance from the lens galaxy, $\bar{\Sigma} (<R)$ 
the average surface density for separations less than $R$ and $\bar{\Sigma}(R)$ 
the average surface density at $R$. If the lens is a satellite galaxy, the 
signal at small $R$ is dominated by the dark matter distribution of the subhalo 
and at large $R$ by the gravitational potential of the host halo. Satellite 
weak lensing studies rely on correctly identifying the group centre to 
accurately determine $r_p$ and also to not accidentally use a central galaxy as 
a lens. In the latter case, the excess surface mass density signal would be 
much larger at small scales due to the large gravitational potential of the 
halo centre. Some studies have used the BHG as the definition of the halo 
centre \citep{Li+14, Sifon+15}. Here, we test the impact of the resulting halo 
centre misidentification on the lensing signal.

We first populate the $z = 0$ Bolshoi-Planck simulation \citep{Riebe+11} with 
galaxies above a luminosity of $10^{9.7} \Lsunh$ according to the 
procedure outline in section \ref{sec:Mock Catalogues}. This luminosity 
threshold roughly corresponds to the threshold used in \cite{Li+14}. We further 
limit this analysis to galaxies living in halos of masses $10^{13} \Msunh < 
M_{\rm vir} < 5 \times 10^{14} \Msunh$. For each galaxy we 
calculate the projected distance $r_p$ to the BHG in the $xy$-plane. Next, we 
follow \cite{Li+14} and bin the galaxies into two bins of projected separation; 
$0.1 < r_p / (\Mpch) < 0.3$ and $0.3 < r_p / (\Mpch) < 0.5$. 
Finally, we use the Bolshoi-Planck particle catalogue 
and the \texttt{delta\_sigma} function provided by \texttt{halotools} to calculate 
the lensing signal around galaxies in each bin of projected separation. We do 
this both for a model with $f_{\rm BNC} = 0$ and for the best-fitting, non-zero 
$f_{\rm BNC}$ model.

In Figure \ref{fig:Satellite_Weak_Lensing} we show the excess surface mass 
density as a function of projected separation from the lens galaxy for all $4$ 
samples. We find that a non-zero $f_{\rm BNC}$ leads to small changes at large 
projected radii from the lens and to a systematically higher excess surface 
mass density at small separations. Thus, not taking a non-zero $f_{\rm BNC}$ 
into account would likely lead to an overestimation of the average subhalo 
mass. However, the effect seems to be mostly independent of the separation 
$r_p$ of the lens galaxy from the BHG. Furthermore, the strength of the effect 
is below current statistical uncertainties. Thus, it seems unlikely that a 
non-zero $f_{\rm BNC}$ would lead to wrong inferences about tidal stripping of 
galaxies \citep{Sifon+15, Li+16, Niemiec+17} or subhalo segregation 
\citep{vdBosch+16}. Our results for the impact of a non-zero $f_{\rm BNC}$ on 
the weak lensing signal of satellite galaxies are in agreement with 
\cite{Li+14}.

\subsection{Impact on Satellite Kinematics}
\label{sec:Satellite_Kinematics}

\begin{figure}
  \centering \includegraphics[width=\columnwidth]{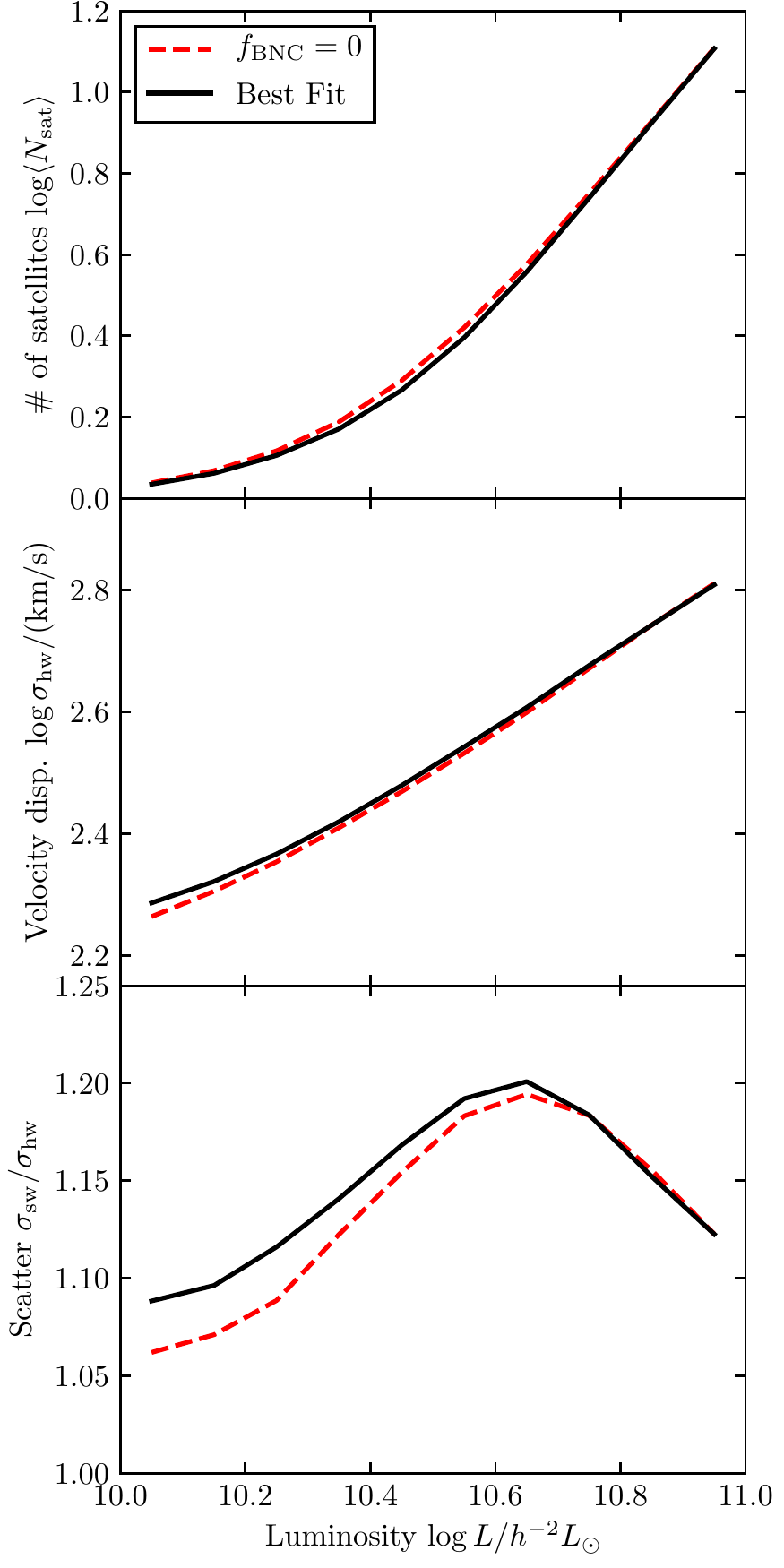} \caption{The
  impact of a non-zero $f_{\rm BNC}$ on the observables in satellite
  kinematic studies similar to \protect\cite{More+11}. In each panel
  we compare the predictions from a model in which all BHGs are
  central (red, dashed) to the best-fitting, non-zero $f_{\rm BNC}$
  model (black, solid). Both models otherwise assume the same CLF
  parameters and were applied to the SMDPL ROCKSTAR halo catalogue. The upper 
  panel shows the average number of secondaries around 
  primaries. The middle panel displays the average
  velocity dispersions as a function of primary luminosity. Finally, the
  lower panel shows the ratio of satellite-weighted to host-weighted
  velocity dispersion, effectively a measure of the scatter in halo
  mass at fixed luminosity. Clearly, a non-zero $f_{\rm BNC}$ similar to that
  inferred here has only a very mild impact on studies of satellite kinematics.}  \label{fig:Satellite_Kinematics}
\end{figure}

Satellite kinematics studies similar to the ones performed
by \cite{vdBosch+04} and \cite{More+11} are a promising avenue to infer
the galaxy--halo connection from small-scale redshift space
distortions \citep[see also e.g.,][]{Prada+03, Conroy+07, Norberg+08,
Li+12, Wojtak+13}. These studies utilize a similar isolation criteria
as used in this study and measure the pairwise velocity dispersion
between primaries and their secondaries as a function of
the luminosity of the primary. The velocity dispersion, in turn, is a measure of
the halo mass. Generally, the latest results of \cite{More+11} are in
good agreement with independent studies using galaxy-galaxy lensing,
group catalogues or galaxy clustering. However, \cite{More+11}
acknowledge that their study seems to lead to systematically higher
values for the inferred average halo masses. Importantly, when
comparing their model to the data, \cite{More+11} assume that the BHG
is always the central. Clearly, if the BHG were not the central, the
pairwise velocity dispersion between the primary and secondaries will be 
higher, simply because satellites have larger
peculiar velocities relative to their halo than centrals.  This
could, in principle, explain part of the discrepancy identified by
\cite{More+11} \citep[see also][]{Dutton+10}. This was previously
discussed in \cite{Skibba+11} and \cite{More+11} who estimated
that this effect could lead to systematic errors in halo mass of up to
a factor of $1.6$. However, to estimate the impact of this effect on
satellite kinematics precisely one needs to know $f_{\rm BNC}$ as a
function of {\it both} halo mass and BHG luminosity. This information was
not available previously, but with the results of this study we can
asses the impact quantitatively.

To study this effect we first populate the \texttt{SMDPL} $z = 0.1$ halo
catalogue with galaxies according to the recipe described in
section $\ref{sec:Mock Catalogues}$ and a luminosity threshold of
$10^{9.5} \Lsunh$. This is the same threshold as used
by \cite{More+11}. We want to explore the effect of a non-zero
$f_{\rm BNC}$ without any observational complications. Thus, we do not
include the projection onto the sphere, the SDSS survey mask,
magnitude thresholds or fibre collisions. Instead, we simply apply our
isolation criteria in the distant observer approximation, i.e., using the 
simulation z-axis as the line-of-sight direction. The
parameters of the cylinder are now $(\Delta V)_\rmh = 5 \sigma$,
$(\Delta V)_\rms = 4000 \kms$, $R_\rms = 0.15 \, \sigma_{200} \Mpch$
and $R_\rmh = 5.33 R_\rms$, the same as in \cite{More+11}. We also
remove secondaries which do not belong to the same halo as the
primaries, and primaries without secondaries. Finally, we measure as
a function of the BHG luminosity the average number of secondaries which 
\cite{More+11} denote by $\langle N_{\rm sat} \rangle$ and the average pairwise 
velocity dispersion between the primary and its secondaries. As discussed
in \cite{More+09a}, the average velocity dispersion can be obtained by
either giving each primary a weight of unity (host-weighted,
$\sigma_{\rm hw}$) or by giving each primary a weight
proportional to the number of secondaries around it
(satellite-weighted, $\sigma_{\rm sw}$). Generally, the
satellite-weighted velocity dispersion estimate is expected to be
higher because primaries with more secondaries tend to
reside in more massive haloes. Thus,
the ratio $\sigma_{\rm sw} / \sigma_{\rm hw}$ is a measure of the
scatter in halo mass for primaries of the same luminosity.

Figure \ref{fig:Satellite_Kinematics} shows the effect of the best-fit
$f_{\rm BNC}$ versus $f_{\rm BNC} = 0$ on these observables. Both
models otherwise use the exact same CLF. In each panel, the red dashed
line shows the result if all BHGs were centrals and the black solid
line the result of our best-fitting model. The upper panel shows the
average number of secondaries as a function of BHG luminosity. As
expected, the number of secondaries scales with the luminosity of
the primary. The best-fitting model predicts a slightly lower
$\langle N_{\rm sat} \rangle$. This is expected because we only assign
galaxies within a certain projected radius around the primary
galaxy. If the primary is a satellite that is far away from
the centre of the halo, the expected number of associations is
smaller. The middle panel shows the impact on the host-weighted
velocity dispersion. Generally, more luminous galaxies have higher
velocity dispersions. As expected, the
pairwise velocity dispersion is higher for our best-fit model compared
to the $f_{\rm BNC} = 0$ case. However, the effect is small, $\sim
3\%$ at most for $\log L / \Lsunh > 10.2$. One reason for this, as we will show below, is that
$f_{\rm BNC}$ as a function of BHG luminosity, averaged over halo
mass, is very low. Additionally, the velocity correlation between satellites also decreases the measured velocity dispersion if a satellite is selected as a primary. The systematic impact on halo mass, using
$M \propto \sigma^3$, is at most $\sim 10\%$. This is below the
statistical uncertainties in \cite{More+11}. Finally, the lower panel
shows the ratio $\sigma_{\rm sw} / \sigma_{\rm hw}$. Our best-fit
model predicts a slightly higher ratio for the same CLF
parameters. This makes sense because at a given BHG luminosity halo
mass and $f_{\rm BNC}$ have a positive correlation. Thus, $\sigma_{\rm
sw}$ being dominated by high-mass haloes is affected more strongly by
a non-zero $f_{\rm BNC}$. In principle, this could lead to an
over-estimate of the scatter in central luminosity as a function of
halo mass if not accounted for. But, again, the effect seems to be too
small to significantly affect the results of \cite{More+11}.

\begin{figure}
  \centering \includegraphics[width=\columnwidth]{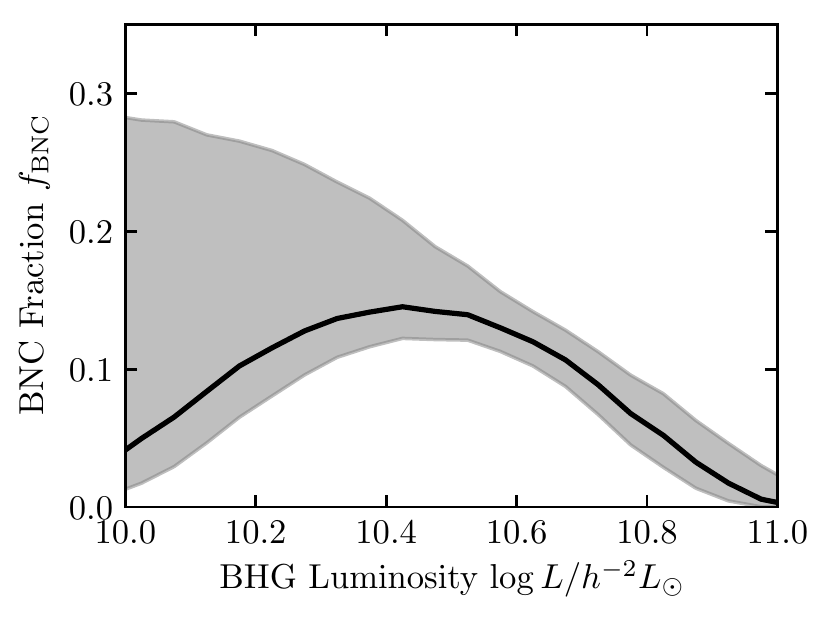} \caption{The
  inferred $f_{\rm BNC}$ as a function of BHG luminosity, marginalized over
  the halo mass function. The solid line represents the median of the posterior 
  while the grey band denotes the $68\%$ uncertainty on the
  posterior prediction.}  \label{fig:BNC_vs_L_2}
\end{figure}

One reason for the low impact of the best-fitting, non-zero
$f_{\rm BNC}$ model on satellite kinematic observables is that $f_{\rm
BNC}$ as a function of BHG luminosity is much lower than as a function
of halo mass. This is shown in Figure \ref{fig:BNC_vs_L_2} where we
show $f_{\rm BNC}$ as a function of BHG luminosity averaged over all
halo masses in SMDPL. The grey band is the $68\%$ posterior of our
default analysis and the solid line the best-fit model. We see that
$f_{\rm BNC}$ has only a weak dependence on BHG luminosity
and does not exceed $\sim 15 \%$. This
might seem at odds with Figure \ref{fig:R_and_S}, which suggests
that $f_{\rm BNC}$ decreases strongly with $L_{\rm
BHG}$. However, for our sample in Figure \ref{fig:R_and_S} we did
require at least $3$ detected secondaries. That means that at the
low-luminosity end this sample is biased towards high-mass haloes,
those with high $f_{\rm BNC}$. Overall, the reason for the low $f_{\rm
BNC}$ is that the halo mass function is relatively steep and that
there is scatter in BHG luminosity at fixed halo mass. That implies
that the number density of galaxies at a given BHG luminosity is
dominated by low-mass haloes with over-luminous BHGs compared to
more massive haloes with under-luminous BHGs. Coincidently, we
find that low-mass haloes with high-luminosity BHGs have very low
$f_{\rm BNC}$. Note also that these conclusions are to some extent
dependent on the extrapolation of our $f_{\rm BNC}$ parametrization to
lower halo masses. This is signified by the large error bars at the
low-luminosity end in Figure \ref{fig:BNC_vs_L_2}. However, we have tested that 
even the $15$ $\{f_0, a_1, a_2, b\}$ combinations from our posterior with the 
lowest likelihood, $\Delta \ln \mathcal{L} \sim 10$ compared to the best-fit 
model, do not change the above conclusions.

We conclude that our results for $f_{\rm BNC}$ imply a very small
impact on satellite kinematic studies \citep{vdBosch+04, More+09b,
More+11, Li+12}. Essentially, assuming $f_{\rm BNC} = 0$ seems to be a
good approximation given current statistical uncertainties.

\subsection{Observed Radial Profile}

\begin{figure}
  \centering
  \includegraphics[width=\columnwidth]{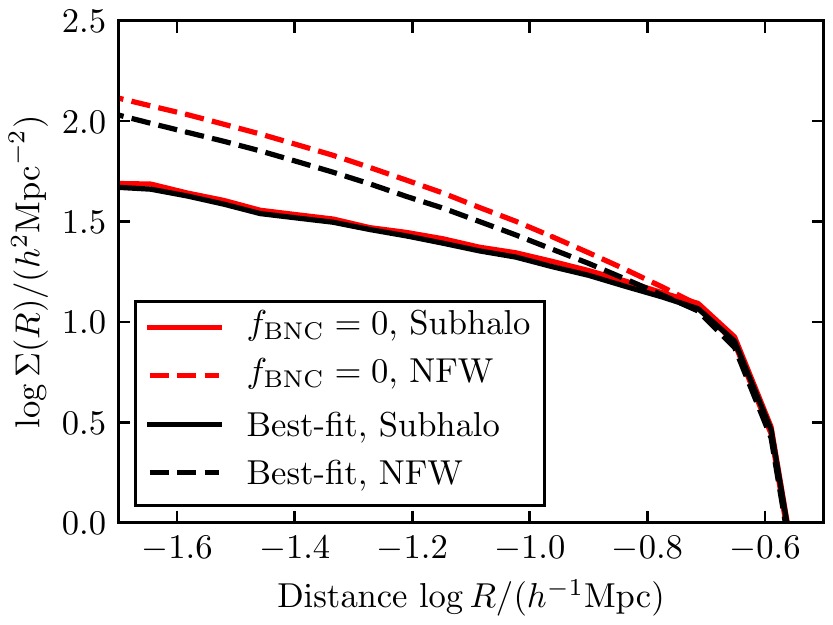}
  \caption{The surface density of secondaries around primaries with $10.5 < 
  \log [L / (h^{-2} L_\odot)] < 10.7$ in mock catalogues. The black lines show 
  results from our best-fit, non-zero $f_{\rm BNC}$ model and the red lines the 
  results for $f_{\rm BNC} = 0$. Conversely, solid lines denote mock catalogues 
  where the satellite phase-space positions are modelled after subhaloes in 
  SMDPL, while dashed lines represent simulations where satellites are 
  assumed to follow an NFW profile. Note that, in both cases, the best-fit model 
  and the model with $f_{\rm BNC} = 0$ yield inferred surface densities that are 
  virtually indistinguishable. Hence, a non-zero $f_{\rm BNC}$ similar to that
  inferred here does not have a significant impact on the inferred surface density profile 
  of satellite galaxies.}
  \label{fig:Surface_Density}
\end{figure}

One crucial assumption going into the modelling of galaxy clustering,
galaxy-galaxy lensing and satellite kinematics is the radial profile
of satellite galaxies \citep[see, e.g.][]{More+09b, Guo+12, Guo+13}. Evidently, the radial profile has a direct
impact on the observed galaxy-galaxy and galaxy-matter
correlation. Furthermore, satellite kinematic studies infer halo
masses by modelling the dynamics using the Jeans equation. This
calculation needs the radial profile of satellite galaxies as an input. The effect of different assumed radial profiles on the
predicted velocity dispersion are considerable. For example, if
subhaloes were anti-biased with respect to the dark matter, the
expected velocity dispersion would be higher by up to $\sim
20\%$ \citep{vdBosch+04}.

\cite{More+09b} derived the radial profile
of satellite galaxies using observations and concluded that satellite
galaxies were anti-biased with respect to the dark matter
distribution. However, this result was derived assuming that all BHGs
were centrals. The concern is that a non-zero $f_{\rm BNC}$ would
systematically alter the observed radial profile. In general, if
some satellite galaxies are BHGs, and consequently mis-classified as
centrals, the inferred radial profile will be less centrally
concentrated. In the case where satellite galaxies are an un-biased
tracer of the underlying mass distribution, this will result in the
incorrect conclusion that satellites are anti-biased with respect to
their dark matter halo.

We address this issue in
Figure \ref{fig:Surface_Density} where we show the measured surface
density of secondaries around primaries with $\log [L /(\Lsunh)] \in
[10.5, 10.7]$. The solid lines are similar to the
ones in Figure \ref{fig:Satellite_Kinematics}, showing the different
observed radial profiles for $f_{\rm BNC} = 0$ and the best-fitting
non-zero $f_{\rm BNC}$ model. In particular, for this model we assume
that satellites follow the subhalo distribution. As in the previous
figure, the effect of interlopers, fibre collisions, projection
effects etc., have been ignored. The down-turn at high radii simply
reflects the radius $R_\rms$ of the cylinder used to associate secondaries with 
primaries. Again, the impact of a non-zero $f_{\rm
BNC}$ is negligible. There is virtually no difference between the
observed profiles. On the other hand, the dashed lines show the
predictions if satellite galaxies would follow an NFW profile with the same 
concentration as the dark matter distribution. For this model
we assigned phase-space coordinates using the \texttt{NFWPhaseSpace}
function of \texttt{halotools}. As shown in
Figure \ref{fig:Surface_Density}, even in the case of a very steep 
radial profile like NFW, the best-fit, non-zero $f_{\rm BNC}$ does not
significantly impact the observed surface number density profile of satellite
galaxies around BHGs.

\section{Conclusion}
\label{sec:Conclusion}

Theoretical models predict that central galaxies of dark matter haloes
should constitute a special population of halo members that are more
luminous than satellite galaxies. Thus, it is commonly assumed that
the BHG of a dark matter halo is the central. In this paper we have
analysed data from the NYU-VAGC \citep{Blanton+05} built upon SDSS
DR7 \citep{Abazajian+09}. In particular, we investigated the
phase-space positions of BHGs with respect to galaxies living in the
same halo. Following \cite{vdBosch+04} and \cite{Skibba+11} we
computed the $\mathcal{R}$ and $\mathcal{S}$ statistic for a sample of galaxies 
with particularly high probabilities of being BHGs. These quantities
measure velocity and spatial offsets, respectively. We compare this to
detailed mock catalogues to model the fraction $f_{\rm BNC}$ of haloes in which 
the BHG is not the central as a function of both
halo mass and BHG luminosity. Compared to \cite{Skibba+11} our model
contains several improvements.
\begin{itemize}

\item Central galaxies are assigned the phase-space positions of halo cores
instead of the bulk velocity of the entire halo. This accounts for the
unrelaxed state of the dark matter distribution of the halo. Since the
effect of phase-space offsets of the central and an increased $f_{\rm
BNC}$ are qualitatively similar \citep{Skibba+11}, this is important
to get an unbiased estimate of $f_{\rm BNC}$.

\item We assign phase-space positions for satellite galaxies using the
positions of resolved subhaloes. Compared to \cite{Skibba+11} this
explicitly models the effects of satellite phase-space correlations
and halo triaxiality.

\item As described in section \ref{subsec:f_BNC_parameters}, our model takes
into account a correlation between satellite occupation and the
probability that the brightest galaxy is not the central. This is important because our sample is biased
towards systems with larger satellite occupation.

\end{itemize}
We model the $\mathcal{R}$ and $\mathcal{S}$ distributions in $4$ bins
of BHG luminosity and $2$ redshift bins. Our best-fit model for
$f_{\rm BNC}$ as a function of halo mass and BHG luminosity is able to
explain the observed $\mathcal{R}$ and $\mathcal{S}$ distributions at
all BHG luminosities ranging from $10^{10}$ to $10^{11.25} \ h^{-2}
L_\odot$ and redshifts ranging from $0.03$ to $0.15$. We find that for
a given halo mass, $f_{\rm BNC}$ is strongly anti-correlated
with the luminosity of the BHG. Such an anti-correlation is expected from
theoretical models of galaxy formation and halo occupation
models like the CLF. This effect can be understood by the fact that it is more 
likely that a satellite exceeds the luminosity of the central if the latter one 
is low. At a fixed BHG luminosity, the probability that
the BHG is not the central is increasing with halo mass. Again, this
outcome is in agreement with predictions from CLF models and naive expectations 
based on the increasing satellite occupation with halo mass. Furthermore,
$f_{\rm BNC}$ marginalized over the BHG luminosity distribution of
each halo increases with halo mass. Comparing our results to previous
studies, we find that $f_{\rm BNC}$ as a function of halo mass is in
good quantitative agreement with the findings of \cite{Skibba+11} and
in slight tension with the results of \cite{Hoshino+15}.

Similar
to \cite{Skibba+11}, we find that our inferred values for $f_{\rm
BNC}$ are surprisingly high compared to the SAMs
of \cite{Croton+06}, \cite{DeLucia+07} and \cite{Monaco+07} and
compared to naive `predictions' based on published CLF models.
We find values for $f_{\rm BNC}$ ranging from $25\%$ to $40\%$
in the halo mass range $10^{13}$ to $10^{15} \ h^{-1} M_\odot$,
whereas the SAMs and CLF generally predict values below
$20\%$. This disagreement could arise from inaccuracies in the SDSS
photometric pipeline \citep{Abazajian+09, Aihara+11} or might signal
an actual short-coming of the theoretical models. For example, it is
possible that SAMs suffer from over-merging or over-quenching of
satellite galaxies. It is also possible that certain common
assumptions of CLF modelling, such as the statistical independence of
central and satellite occupation, are violated. It will be
interesting to extend our analysis to lower luminosities and halo masses,
which may be possible with surveys such as GAMA \citep[][]{Driver+11}.

Using our best-fit model for $f_{\rm BNC}$ we explore the impact of a
non-zero $f_{\rm BNC}$ onto studies that implicitly assume that the
BHG is the central. Satellite kinematic
studies \citep[e.g.,][]{vdBosch+04, Conroy+07, Norberg+08, More+09a,
More+09b, More+11, Li+12} infer the average halo mass of an ensemble
of BHGs by modelling the redshift-space correlations with other halo
members. Typically, in order to model the inferred halo masses
it is explicitly assumed that the BHG is always the central,
i.e. $f_{\rm BNC} = 0$. Here, we have shown that this simplification
only leads to a very modest bias in the inferred halo
masses of order $\lesssim 10\%$. We also explored the impact of
assuming $f_{\rm BNC} = 0$ on the observed radial profile of satellite
galaxies around BHGs. Once again, we find that the impact is subtle
at most. Most importantly, it is unlikely that the inference that
satellite galaxies are strongly anti-biased (i.e., less centrally
concentrated) with respect to the dark matter
distribution \citep[e.g.,][]{Yang+05, Chen08, More+09b}, is solely caused by
this assumption. Finally, our best-fit models allow to test the impact
of this assumption for other observables such as
galaxy-clustering and galaxy-galaxy lensing. Many of the findings of the 
present analysis will also be used in a future work employing satellite 
kinematics and gravitational lensing.

\section*{Acknowledgements}

This research was supported in part by the National Science Foundation
under Grant No. NSF PHY-1125915. JUL was supported by a KITP graduate
fellowship. FvdB and JUL are supported by the US National Science
Foundation through grant AST 1516962. APH is funded through the Yale Center for 
Astronomy \& Astrophysics Prize fellowship. ARZ is funded by the Pittsburgh 
Particle physics Astrophysics and Cosmology Center (Pitt PACC) at the Univer-
sity of Pittsburgh and by the NSF through grant NSF AST 1517563. YYM is 
supported by the Samuel P.\ Langley PITT PACC Postdoctoral Fellowship. This 
work was supported by the HPC facilities operated by, and the staff of, the 
Yale Center for Research Computing.

This work made use of the following software
packages: \texttt{matplotlib} \citep{Hunter07}, \texttt{SciPy}, \texttt{NumPy} \citep{vdWalt+11}, \texttt{Astropy} \citep{Astropy13}, \texttt{Cython} \citep{Behnel+11}, \texttt{halotools} \citep{Hearin+16}, \texttt{Corner} \citep{Foreman-Mackey+16}, \texttt{MultiNest} \citep{Feroz+08,
Feroz+09,
Feroz+13}, \texttt{PyMultiNest} \citep{Buchner+14}, \texttt{mangle} 
\citep{Hamilton+04, Swanson+08}
and \texttt{pymangle}\footnote{\url{https://github.com/esheldon/pymangle}}. We 
are very grateful to all contributors to the
above mentioned software packages which helped to greatly expedite
this work. We thank the anonymous referee for an insightful report that 
significantly improved the presentation of this paper. We are also grateful to 
Alexie Leauthaud, Surhud More, Sarah Brough and Marina Trevisan for their helpful discussions.

% http://sdss.physics.nyu.edu/vagc/cites.html
Funding for the Sloan Digital Sky Survey (SDSS) has been provided by
the Alfred P. Sloan Foundation, the Participating Institutions, the
National Aeronautics and Space Administration, the National Science
Foundation, the U.S. Department of Energy, the Japanese
Monbukagakusho, and the Max Planck Society. The SDSS Web site is
http://www.sdss.org/.

The SDSS is managed by the Astrophysical Research Consortium (ARC) for
the Participating Institutions. The Participating Institutions are The
University of Chicago, Fermilab, the Institute for Advanced Study, the
Japan Participation Group, The Johns Hopkins University, Los Alamos
National Laboratory, the Max-Planck-Institute for Astronomy (MPIA),
the Max-Planck-Institute for Astrophysics (MPA), New Mexico State
University, University of Pittsburgh, Princeton University, the United
States Naval Observatory, and the University of Washington.

% http://gavo.mpa-garching.mpg.de/Millennium/Help?page=credits
The Millennium Simulation databases used in this paper and the web
application providing online access to them were constructed as part
of the activities of the German Astrophysical Virtual Observatory
(GAVO).

% https://www.cosmosim.org/cms/simulations/multidark-bolshoi-project/
The authors gratefully acknowledge the Gauss Centre for Supercomputing
e.V. (www.gauss-centre.eu) and the Partnership for Advanced
Supercomputing in Europe (PRACE, www.prace-ri.eu) for funding the
MultiDark simulation project by providing computing time on the GCS
Supercomputer SuperMUC at Leibniz Supercomputing Centre (LRZ,
www.lrz.de).
The Bolshoi simulations have been performed within the Bolshoi project
of the University of California High-Performance AstroComputing Center
(UC-HiPACC) and were run at the NASA Ames Research Center.

\label{lastpage}

\bibliographystyle{mnras}
\bibliography{bibliography}

\section*{Appendix A: CLF Predictions}
\label{sec:Appendix_A}

Here, we describe several relations regarding the probability of the BHG to be 
a satellite in CLF models. A common assumption in CLF modelling is that each 
halo hosts exactly one central (at least for the mass scale of interest here). 
This requires
\begin{equation}
\int\limits_0^\infty \Phi_\rmc (L_\rmc | M) dL_\rmc = 1.
\end{equation}
Particularly, this implies that among those systems that do not have a central
galaxy brighter than $L_{\rm max}$, their central CLF $\Phi_\rmc^\star(L_\rmc | M, 
L_{\rm max})$ must obey
\begin{equation}
\int\limits_0^{L_{\rm max}} \Phi_\rmc^\star(L_\rmc | M, L_{\rm max}) dL_\rmc = 
1,
\end{equation}
from which we find
\begin{equation}
\Phi_\rmc^\star(L_\rmc | M, L_{\rm max}) = \frac{\Phi_\rmc(L_\rmc | 
M)}{\int\limits_0^{L_{\rm max}} \Phi_\rmc (L_\rmc | M) dL_\rmc}.
\end{equation}
On the other hand, the occupation of satellites is assumed to follow a Poisson 
distribution and that luminosities of satellites are uncorrelated. This implies
\begin{equation}
\Phi_\rms^\star (L_\rms | M, L_{\rm max}) = \Phi_\rms (L_\rms | M).
\end{equation}
Now, given a halo with no central and satellite above a luminosity $L_{\rm 
BHG}$ but a galaxy with luminosity $L \in [L_{\rm BHG} - dL, L_{\rm BHG}]$, the 
probability that this galaxy is a satellite is
\begin{multline}
f_{\rm BNC}(L_{\rm BHG}, M) = \frac{\Phi_\rms (L_{\rm BHG} | 
M)}{\Phi_\rmc^\star(L_{\rm BHG} | M, L_{\rm BHG}) + \Phi_\rms (L_{\rm BHG} | 
M)},
\end{multline}
from which Eq.~(\ref{eq:f_BNC_CLF}) follows.

The probability that a satellite brighter than some 
threshold luminosity $L_{\rm th}$ is brighter than the central with luminosity 
$L_\rmc$ is given by
\begin{equation}
p(L_\rms > L_\rmc | L_\rms > L_{\rm th}) = \frac{\int\limits_{L_\rmc}^\infty 
\Phi_\rms (L_\rms | M) dL_\rms}{\int\limits_{L_{\rm th}}^\infty \Phi_\rms 
(L_\rms | M) dL_\rms} = \frac{\langle n_{\rm BTC} \rangle}{\langle 
n_{\rm sat} \rangle}
\end{equation}
Let us assume that the number of satellites fainter than $L_{\rm c}$ is fixed, 
$n_{\rm sat} - n_{\rm BTC} = n_{\rm FTC}$. Then the probability to have $n_{\rm 
BTC}$ satellites brighter than $L_\rmc$ is
\begin{align}
p(n_{\rm BTC} | n_{\rm FTC}) =& p(n_{\rm sat} = n_{\rm FTC} + n_{\rm BTC}) 
\times \nonumber\\
& p(n_{\rm BTC} | n_{\rm sat} = n_{\rm FTC} + n_{\rm BTC}) \nonumber\\
\propto& \frac{\langle n_{\rm sat} \rangle^{n_{\rm BTC}}}{(n_{\rm FTC} + n_{\rm 
BTC})!} \genfrac(){0pt}{}{n_{\rm FTC} + n_{\rm BTC}}{n_{\rm BTC}} \times 
\nonumber\\
&\left( \frac{\langle n_{\rm BTC} \rangle}{\langle n_{\rm sat} \rangle} 
\right)^{n_{\rm BTC}} \nonumber\\
\propto& \frac{\langle n_{\rm BTC} \rangle^{n_{\rm BTC}}}{n_{\rm BTC}!}.
\end{align}
That means that $n_{\rm BTC}$ also follows a Poisson distribution and is 
completely independent of the occupation with satellites fainter than $L_\rmc$. 
From this, we can easily show that the probability that any satellite is 
brighter than the central luminosity $L_\rmc$ is
\begin{align}
f_{\rm BNC} =& \sum\limits_{n_{\rm BTC} = 1}^\infty \frac{\langle n_{\rm BTC} 
\rangle^{n_{\rm BTC}}}{n_{\rm BTC}!} \exp \left( -\langle n_{\rm BTC} \rangle 
\right) \nonumber\\
=& 1 - \exp \left( -\langle n_{\rm BTC} \rangle \right).
\end{align}
Finally, the probability that a halo of mass $M$ has no galaxy above luminosity 
$L_{\rm max}$ is
\begin{align}
p(L_{\rm BHG} \leq L_{\rm max}) =& \left( \int\limits_0^{L_{\rm max}} 
\Phi_\rmc(L_\rmc | M) dL_\rmc \right) \times \nonumber\\
& \exp\left[- \int\limits_{L_{\rm max}}^\infty \Phi_\rms (L_\rms | M) 
dL_\rms \right].
\end{align}
The first term in the right-hand side of the above equation describes the 
probability that the central is not brighter than $L_{\rm max}$, whereas the 
second term describes the probability that no satellite galaxy is brighter than 
$L_{\rm max}$. Ultimately, the CLF of the BHG follows
\begin{align}
\Phi_{\rm BHG} (L | M) =& p(L \leq L_{\rm BHG}) (\Phi_\rmc^\star (L | M, L_{\rm 
BHG}) + \Phi_\rms (L | M)) \nonumber\\
=& \left[ \Phi_\rmc(L | M) + \Phi_\rms(L | M) \int\limits_0^{L_{\rm BHG}} 
\Phi_\rmc(L_\rmc | M) dL_\rmc \right] \nonumber\\
&\times \exp\left[- \int\limits_{L_{\rm BHG}}^\infty \Phi_\rms 
(L_\rms | M) dL_\rms \right].
\end{align}
\end{document}